\pdfoutput=1
\documentclass[12pt,a4paper]{article}

\usepackage{natbib}

\usepackage{ucs}
\usepackage[utf8x]{inputenc}
\usepackage[T1]{fontenc}

\usepackage{geometry}
\usepackage{fullpage}

\usepackage{xspace}
\usepackage{enumitem}
\setlist{nosep}

\usepackage{amsmath}
\usepackage{amsthm}
\usepackage{stmaryrd} 
\usepackage{stix} 

\usepackage{tikz}  
\usetikzlibrary{intersections}
\usetikzlibrary{calc}
\usetikzlibrary{patterns}
\usetikzlibrary{decorations.markings}
\usetikzlibrary{decorations.pathmorphing}
\usetikzlibrary{decorations.pathreplacing}

\usepackage{hyperref}

\usepackage{subcaption}

\graphicspath{{./Pictures/}}

\usepackage{h_general}
\usepackage{h_vocabulary}
\usepackage{h_graphics}

\usepackage{Add}
\usepackage{comment}

\newcommand{\TilePolice}[1]{\textsf{#1}\xspace}

\newcommand{\GluePolice}[1]{\textsf{#1}\xspace}

\newcommand{\Tile}[3][last tile]{%
  \path (#2) \CoorNode{#1} ;
  \draw (#2) node[inner sep=0cm,minimum size=.8\unitlength,draw] (#1) {\TilePolice{#3}} ;
}

\newcommand{\LinkTile}[6]{%
  \draw (#4) --
  node[sloped,pos=.15,above,inner sep=0.1em] {\tiny #1}
  node[sloped,pos=.85,inner sep=0.1em,above] {\tiny #2}
  node[#6,inner sep=0.2em] {\scriptsize \GluePolice{#3}}
  (#5) ;%
}

\newcommand{\LinkBotTop}[3][]{%
  \LinkTile{}{}{#1}{#2}{#3}{left}%
}

\newcommand{\LinkLeRi}[3][]{%
  \LinkTile{}{}{#1}{#2}{#3}{above}%
}

\newcommand{\GlueWidth}{.25}

\newcommand{\GlueTop}[2][]{%
  \begin{scope}[shift={(#2)}]
    \begin{scope}[shift={(0,.5)}]
      \draw (0,0) node[fill=LightBrown,inner sep=0cm,minimum width=2*\GlueWidth\unitlength,minimum height=.2\unitlength,draw] {\tiny \GluePolice{#1}} ;
    \end{scope}
  \end{scope}
}

\newcommand{\GlueRi}[2][]{%
  \begin{scope}[shift={(#2)}]
    \begin{scope}[shift={(.5,0)}]
      \draw (0,0) node[fill=LightBrown,inner sep=0cm,minimum height=2*\GlueWidth\unitlength,minimum width=.2\unitlength,draw] {\tiny \GluePolice{#1}} ;
    \end{scope}
  \end{scope}
}

\newcommand{\SEED}[1]{%
  \begin{scope}[shift={(#1)}]
    \newcommand{\Rad}{.325}
    \draw[ultra thick,densely dotted] (\Rad,\Rad) -- (\Rad,-\Rad) -- (-\Rad,-\Rad) -- (-\Rad,\Rad) -- cycle;
  \end{scope}
}

\newcommand{\GlueLe}[2][]{%
  \begin{scope}[shift={(#2)}]
    \begin{scope}[shift={(-1,0)}]
      \GlueRi[#1]{0,0}
    \end{scope}
  \end{scope}
}

\newcommand{\GlueBot}[2][]{%
  \begin{scope}[shift={(#2)}]
    \begin{scope}[shift={(0,-1)}]
      \GlueTop[#1]{0,0}
    \end{scope}
  \end{scope}
}

\pgfdeclarepatternformonly{NE lines}{\pgfqpoint{-1pt}{-1pt}}{\pgfqpoint{8pt}{8pt}}{\pgfqpoint{6pt}{6pt}}%
{
  \pgfsetlinewidth{0.4pt}
  \pgfpathmoveto{\pgfqpoint{0pt}{0pt}}
  \pgfpathlineto{\pgfqpoint{6.2pt}{6.2pt}}
  \pgfusepath{stroke}
}

\pgfdeclarepatternformonly{NW lines}{\pgfqpoint{-1pt}{-1pt}}{\pgfqpoint{8pt}{8pt}}{\pgfqpoint{6pt}{6pt}}%
{
  \pgfsetlinewidth{0.4pt}
  \pgfpathmoveto{\pgfqpoint{0pt}{6pt}}
  \pgfpathlineto{\pgfqpoint{6.2pt}{-0.2pt}}
  \pgfusepath{stroke}
}

\makeatletter
\def\IfNextChar#1#2#3#4{%
  \ifx#1#4\@empty\@empty
  \expandafter\@firstoftwo
  \else
  \expandafter\@secondoftwo
  \fi
  {#2}{#3}%
}
\makeatother

\def\IfPathWest#1{%
  \IfNextChar{W}{ - - ++ (-1,0)\IfPathSouth}{%
  }{#1}%
}%
\def\IfPathEast#1{%
  \IfNextChar{E}{ - - ++ (1,0)\IfPathSouth}{%
    \IfPathWest#1
  }{#1}%
}%
\def\IfPathNorth#1{%
  \IfNextChar{N}{ - - ++ (0,1)\IfPathSouth}{%
    \IfPathEast#1
  }{#1}%
}%
\def\IfPathSouth#1{%
  \IfNextChar{S}{ - - ++ (0,-1)\IfPathSouth}{%
    \IfPathNorth #1
  }{#1}%
}%

\newcommand{\PreparePath}[2]{%
  \expandafter\edef\csname Path#1\endcsname{\IfPathSouth #2.}
}

\newcommand{\TransientPathStyleAdd}[1]{%
  \tikzstyle{TransientPathStyleAddStyle}=[#1]}
\TransientPathStyleAdd{}

\newcommand{\PeriodicPathStyleAdd}[1]{%
  \tikzstyle{PeriodicPathStyleAddStyle}=[#1]}
\PeriodicPathStyleAdd{}

\JDLvocabulary{\RealSet}{\JDLvocabularyMathBBXspace{R}}{}{Set of all real numbers}
\JDLvocabulary{\RationalSet}{\JDLvocabularyMathBBXspace{Q}}{}{Set of all rational numbers}
\JDLvocabulary{\IntegerSet}{\JDLvocabularyMathBBXspace{Z}}{}{Set of all integers}
\JDLvocabulary{\NaturalSet}{\JDLvocabularyMathBBXspace{N}}{}{Set of all natural numbers}
\providecommand{\IntegerInterval}[2]{\JDLvocabularyMathXspace{{\llbracket}#1,#2{\rrbracket}}}

\JDLvocabulary{\DomainFunction}{\JDLvocabularyMathXspace{\mathsf{dom}}}{}{Domain of a path, i.e. visited vertices}
\newcommand{\Domain}[1]{\JDLvocabularyMathXspace{\DomainFunction(#1)}}

\JDLvocabulary{\BoundaryFunction}{\JDLvocabularyMathXspace{\partial}}{}{Boundary of a region/subgraph of  $\IntegerSet^2$}
\newcommand{\Boundary}[1][]{\JDLvocabularyMathXspace{\BoundaryFunction #1}}
\JDLvocabulary{\InsideFunction}{\JDLvocabularyMathXspace{\mathsf{inside}}}{}{Inside of a sub set of $\IntegerSet^2$}

\JDLvocabulary{\SubGraphOf}{\JDLvocabularyMathXspace{\sqsubseteq}}{}{Sub-graph inclusion (vertices and edges)}
\JDLvocabulary{\SubGraphCup}{\JDLvocabularyMathXspace{\sqcup}}{}{Union of sub-graphs (vertices and edges)}
\JDLvocabulary{\SubGraphCap}{\JDLvocabularyMathXspace{\sqcap}}{}{Intersection of sub-graphs (vertices and edges)}

\JDLvocabulary{\CupDis}{\ensuremath{\uplus}}{}{Disjoint union}
\JDLvocabulary{\CupSym}{\ensuremath{\triangle}}{}{Symmetric difference}

\DeclareRobustCommand{\Vect}[1]{\JDLvocabularyMathXspace{\overrightarrow{#1}}}
\newcommand{\Backward}[1]{\JDLvocabularyMathXspace{\overline{#1}}}

\newcommand{\Reverse}[1]{%
  \begin{tikzpicture}%
    \path (0,0) node[inner sep=0em,text depth=0] (RB) {\ensuremath{#1}} ;
     \draw[solid,black,semithick,<<-] ([yshift=.2em]RB.north west) -- ([yshift=.2em]RB.north east) ;
  \end{tikzpicture}%
  \xspace%
}

\newcommand{\FreePath}[1]{%
  \JDLvocabularyMathXspace{%
    \ifJDLvocabularyTrackLocal%
    \colorlet{current}{.}%
    \color{Blue}%
    \JDLvocabularyTrackLocalfalse
    \left\langle%
      \JDLvocabularyTrackLocaltrue%
      \color{current}%
      \else%
      \left\langle
        \fi%
        #1%
      \ifJDLvocabularyTrackLocal%
      \colorlet{current}{.}%
      \color{Blue}%
      \JDLvocabularyTrackLocalfalse
      \right\rangle
      \JDLvocabularyTrackLocaltrue%
      \color{current}%
      \else%
    \right\rangle
    \fi%
  }%
}

\JDLvocabulary{\ZZOrigin}{\JDLvocabularyTextXspace{{\rm\textbf{o}}}}{}{The origin of $\IntegerSet^2$, $=(0,0)$}

\JDLvocabulary{\East}{\JDLvocabularyTextXspace{{\rm\textbf{e}}}}{}{Vector for East, $=(1,0)$}
\JDLvocabulary{\North}{\JDLvocabularyTextXspace{{\rm\textbf{n}}}}{}{Vector for North, $=(0,1)$}
\JDLvocabulary{\South}{\JDLvocabularyTextXspace{{\rm\textbf{s}}}}{}{Vector for South, $=(0,-1)$}
\JDLvocabulary{\West}{\JDLvocabularyTextXspace{{\rm\textbf{w}}}}{}{Vector for West, $=(-1,0)$}

\JDLvocabulary{\EmptyWord}{\JDLvocabularyMathXspace{\varepsilon}}{}{Empty word / empty free path}

\JDLvocabulary{\RotateFunction}{\JDLvocabularyMathXspace{\mathsf{rot}}}{}{Rotations of a finite free path}
\newcommand{\Rotate}[1]{\JDLvocabularyMathXspace{\RotateFunction(#1)}}

\JDLvocabulary{\TAS}{\JDLvocabularyMathCalXspace{T}}{}{Tile Assembly System}
\JDLvocabulary{\GlueSet}{\JDLvocabularyMathXspace{\Sigma}}{}{Set of glues}
\JDLvocabulary{\TileSet}{\JDLvocabularyMathXspace{T}}{}{Set of tile types}
\JDLvocabulary{\Assembly}{\JDLvocabularyMathXspace{\alpha}}{}{Some assembly of \TAS}
\JDLvocabulary{\AssemblyOther}{\JDLvocabularyMathXspace{\beta}}{}{Some other assembly of \TAS}
\JDLvocabulary{\Path}{\JDLvocabularyMathXspace{\pi}}{}{Some path in $\IntegerSet^2$}
\JDLvocabulary{\PathOther}{\JDLvocabularyMathXspace{\tau}}{}{Some other path in $\IntegerSet^2$}

\JDLvocabulary{\DirectionSet}{\JDLvocabularyMathXspace{D}}{}{The set of directions, $\{\East,\North,\South,\West\}$.}

\JDLvocabulary{\IMinPath}{\JDLvocabularyMathXspace{p}}{}{Any Infinite minimal path in the assembly starting on the seed}
\JDLvocabulary{\IMinPathSuf}{\JDLvocabularyMathXspace{\hat{p}}}{}{1st Suffix of $p$ where all glues and tiles used appear infinitely often}

\JDLvocabulary{\VIMinPath}{\JDLvocabularyMathXspace{V_{\IMinPath}}}{}{Tiles infinitely used by an infinite path p}
\JDLvocabulary{\EIMinPath}{\JDLvocabularyMathXspace{E_{\IMinPath}}}{}{Oriented glues infinitely used by an infinite path p}
\JDLvocabulary{\GIMinPath}{\JDLvocabularyMathXspace{G_{\IMinPath}}}{}{Oriented graph corresponding to the infinitely repeated part of an infinite path p}

\JDLvocabulary{\VIMinPathSuf}{\JDLvocabularyMathXspace{V_{\IMinPathSuf}}}{}{Tiles infinitely used by an infinite path p}
\JDLvocabulary{\EIMinPathSuf}{\JDLvocabularyMathXspace{E_{\IMinPathSuf}}}{}{Oriented glues infinitely used by an infinite path p}
\JDLvocabulary{\GIMinPathSuf}{\JDLvocabularyMathXspace{G_{\IMinPathSuf}}}{}{Oriented graph corresponding to the infinitely repeated part of an infinite path p}

\JDLvocabulary{\IMinPathCycle}{\JDLvocabularyMathXspace{c}}{}{Cycle infinitely repeated in p}

\JDLvocabulary{\NonCausalFunction}{\JDLvocabularyMathXspace{\mathsf{non\_causal}}}{}{Non-causal vertices for a vertex in $\IntegerSet^2$}
\newcommand{\NonCausal}[1]{\JDLvocabularyMathXspace{\NonCausalFunction(#1)}}

\JDLvocabulary{\CoGrowFunction}{\JDLvocabularyMathXspace{\mathsf{coGrow}}}{}{(right) co-growth of a free path inside 2 regions}
\newcommand{\CoGrow}[4]{\JDLvocabularyMathXspace{\CoGrowFunction\left(\,{#1},\,{#2},\,{#3},\,{#4}\,\right)}}
\JDLvocabulary{\LCoGrowFunction}{\JDLvocabularyMathXspace{\mathsf{l{-}coGrow}}}{}{(left) Co-growth of a free path inside 2 regions}
\newcommand{\LCoGrow}[4]{\JDLvocabularyMathXspace{\LCoGrowFunction\left(\,{#1},\,{#2},\,{#3},\,{#4}\,\right)}}

\JDLvocabulary{\CoGrowStart}{\JDLvocabularyMathXspace{\circledcirc}}{}{Co-growth starting tile}

\JDLvocabulary{\GridFunction}{\JDLvocabularyMathXspace{\mathsf{grid}}}{}{Subgraph of $\IntegerSet^2$ forming a grid}

\JDLvocabulary{\CombFunction}{\JDLvocabularyMathXspace{\mathsf{comb}}}{}{Subgraph of $\IntegerSet^2$ forming a comb}

\JDLvocabulary{\AlphaMax}{\JDLvocabularyMathXspace{\alpha_{max}}}{}{The unique maximal assembly of \TAS}
\JDLvocabulary{\Seed}{\JDLvocabularyMathXspace{\mathsf{\sigma}}}{}{The seed tile type}

\JDLvocabulary{\AnyTile}{\JDLvocabularyMathXspace{\mathsf{t}}}{}{Any tile type in $T$}

\JDLvocabulary{\LineSEast}{\JDLvocabularyMathXspace{\ZZOrigin\East}}{}{The (one-side) infinite line $(0,0).\East^{\omega}$}
\JDLvocabulary{\LineSWest}{\JDLvocabularyMathXspace{\ZZOrigin\West}}{}{The (one-side) infinite line $(0,0).\West^{\omega}$}

\JDLvocabulary{\PhasedPeriod}{\JDLvocabularyMathXspace{q}}{}{Phase in a periodic free path}

\JDLvocabulary{\Quipu}{\JDLvocabularyMathXspace{Q}}{}{Some quipu}
\JDLvocabulary{\QuipuZero}{\JDLvocabularyMathXspace{\Quipu_{0}}}{}{Initial quipu for a filtration}
\newcommand{\QuipuI}{\JDLvocabularyMathXspace{\Quipu_{i}}}

\JDLvocabulary{\QuipuOther}{\JDLvocabularyMathXspace{\Quipu'}}{}{Some other quipu}
\JDLvocabulary{\QuipuLim}{\JDLvocabularyMathXspace{\Quipu_{\infty}}}{}{Infinite tree-quipu in the limit}
\JDLvocabulary{\Root}{\JDLvocabularyMathXspace{r}}{}{Root vertex of a quipu}
\JDLvocabulary{\VertexLabel}{\JDLvocabularyMathXspace{\eta}}{}{Vertex labelling with tile types of a quipu}
\JDLvocabulary{\ArcLabel}{\JDLvocabularyMathXspace{\lambda}}{}{Arc labelling with directions for a quipu}
\JDLvocabulary{\AlphaQuipu}{\JDLvocabularyMathXspace{\alpha_{\Quipu}}}{}{Assembly generated by a quipu \Quipu}
\JDLvocabulary{\AlphaQuipuOther}{\JDLvocabularyMathXspace{\alpha_{\QuipuOther}}}{}{Assembly generated by a quipu \QuipuOther}
\JDLvocabulary{\AlphaQuipuZero}{\JDLvocabularyMathXspace{\alpha_{\QuipuZero}}}{}{Assembly generated by \QuipuZero}
\JDLvocabulary{\AlphaQuipuI}{\JDLvocabularyMathXspace{\alpha_{\QuipuI}}}{}{Assembly generated by $\Quipu_{i}$}
\JDLvocabulary{\AlphaQuipuLim}{\JDLvocabularyMathXspace{\alpha_{\QuipuLim}}}{}{Assembly generated by \QuipuLim}

\JDLvocabulary{\PseudoComb}{\JDLvocabularyMathXspace{\mu}}{}{Pseudo-comb}
\JDLvocabulary{\PseudoCombOther}{\JDLvocabularyMathXspace{\mu'}}{}{Pseudo-comb}

\JDLvocabulary{\CoverFunction}{\JDLvocabularyMathXspace{\mathsf{cover}}}{}{Part of $\IntegerSet^2$ covered by a quipu vertex}

\JDLvocabulary{\ZoneZero}{\JDLvocabularyMathXspace{Z_0}}{}{Finite zone around the seed}
\JDLvocabulary{\ZoneOne}{\JDLvocabularyMathXspace{Z_1}}{}{Backbone zone}
\JDLvocabulary{\ZoneTwo}{\JDLvocabularyMathXspace{Z_2}}{}{Proto-teeth zone}

\JDLvocabulary{\SetCombRepresentative}{\JDLvocabularyMathCalXspace{S}}{}{set of comb representatives}

\newcommand{\PathTo}[2]{\JDLvocabularyMathXspace{#1_{^{*\!}{#2}}}}
\newcommand{\PathBefore}[2]{\JDLvocabularyMathXspace{#1_{^{+\!}{#2}}}}
\newcommand{\PathFrom}[2]{\JDLvocabularyMathXspace{#1_{{#2}^{\!*}}}}
\newcommand{\PathAfter}[2]{\JDLvocabularyMathXspace{#1_{{#2}^{\!+}}}}

\JDLvocabularySymbol{\PathTo{\Path}{i}}{Path up to  index $i$, included. It correspond to $\Path_{\IntegerInterval{0}{i}}$}
\JDLvocabularySymbol{\PathBefore{\Path}{i}}{Path up to  index $i$, excluded. It correspond to $\Path_{\IntegerInterval{0}{i-1}}$}
\JDLvocabularySymbol{\PathFrom{\Path}{i}}{Path from index $i$, included. It correspond to $\Path_{\IntegerInterval{i}{+\infty}}$}
\JDLvocabularySymbol{\PathAfter{\Path}{i}}{Path from index $i$, excluded. It correspond to $\Path_{\IntegerInterval{i+1}{+\infty}}$}

\JDLvocabulary{\FreePathPath}{\ensuremath{\FreePath{\Path}}}{}{Free path associated to \Path}

\begin{document}


\title{Deterministic Non-cooperative Binding in Two-Dimensional Tile Assembly Systems Must Have Ultimately Periodic Paths}

\author{%
  J{\'e}r{\^o}me \textsc{Durand-Lose}\footnote{Universit\'e d'Orl\'eans, INSA Centre-Val-de-Loire, LIFO \'EA 4022, F-45067 Orl\'eans, France
    and LIX, UMR 7161, CNRS, Ecole Polytechnique, Palaiseau, France%
  }
  \and Hendrik Jan \textsc{Hoogeboom}
  \footnote{Leiden University, The Netherlands%
  }
  \and Nata\v sa \textsc{Jonoska} \footnote{University of South Florida, Department of Mathematics and Statistics, Tampa FL, 33620, USA%
  }
}

\date{\today}

\maketitle

\begin{abstract}
  We consider non-cooperative binding, so-called `temperature 1', in  deterministic or directed (called here {\it confluent}) tile self-assembly systems in two dimensions and show a necessary and sufficient condition for such system to have an ultimately periodic assembly path.
  We prove that an infinite maximal assembly has an ultimately periodic assembly path if and only if it contains an infinite assembly path that does not intersect a periodic path in the $\IntegerSet^2$ grid.
  Moreover we show that every infinite assembly must satisfy this condition, and therefore, contains an ultimately periodic path.
  This result is obtained through a superposition and a combination of two paths that produce a new path with desired properties, a technique that we call {\it co-grow} of two paths.
  
  The paper is an updated and improved version of the first part of \cite{durand-lose+hoogeboom+jonoska19arxiv}.
\end{abstract}

\renewcommand{\abstractname}{Keywords}

\begin{abstract}
  Tile assembly system; Directed (confluent) system; Non-cooperation; Ultimately periodic. 
\end{abstract}


\section{Introduction}
\label{sec:intro}

The abstract tile self-assembly model (aTAM) was introduced by Winfree in 1998 \citep{Winf98} as a theoretical model that describes DX DNA self-assembly processes.
The DX molecule can be designed with four sticky ends such that their assembly forms a 2D surface area \citep{Ferong} as if tiled with square tiles.
Hence, motivated by Wang tiles, the abstract tile assembly model is based on square tiles with colored edges (`glues', simulating the DNA sticky ends).
Starting from a seed assembly (or a seed tile) the assembly grows through matching glue attachments of tiles.
Unlike Wang tiles, the glues have strength and when the matching glues are strong enough, the tiles can attach to the growing structure, although there may be mismatched glues on other sides of the tile.
It was observed that two or three weaker matching glues can achieve bonding of the tile similar to the bonding with a higher strength single glue.
The notion of glue strength is captured in the model through a parameter called `temperature'.
If all tiles have uniform strength sticky ends allowing attachment, then it is said that the model describes `temperature~1' bonding. When there are two or more strengths on the sticky ends, the temperature can be higher than one.
In temperature 2, for example, a tile can attach to the growing assembly either by matching of a single glue of strength 2, or, at least two weaker tile glues of strength one are matched.
The latter case is called `cooperative' binding.
Such bond cooperation is not needed (although can appear) when tiles have uniform strength on their sticky ends, or when the model runs at `temperature 1', or non-cooperative binding.

There are several experimental assemblies of DNA-based tile arrays that show that aTAM can carry out computation.
These include a binary counter using four DX-based tiles \citep{Evans2015}, binary addition by TX molecules \citep{LaBean2000}, Sierpinski triangle as a pattern on a substrate \citep{Sierpinski,Fujibayashi2008}, transducer simulations by TX molecules \citep{Banani}, and the most recent one where different combinations of input tiles achieve a variety of computations \citep{Damien-Nature-2019}.

In his thesis \citep{Winf98}, Winfree showed that the abstract tile assembly model at temperature 2 can assemble (simulate) a trace of computation of any Turing machine, thereby proving that the model has universal computational power.
At the same time it was conjectured that temperature~1 systems have strictly lower computational power. 
In \citep{rothemund+winfree00stoc}, it was also observed that temperature~2 aTAM can assemble certain structures, such as squares, with much smaller number of tile types compared with temperature~1 systems, indicating a possible difference in the computational power of the two models. 

The theoretical proofs for universal computational power of aTAM in temperature 2, as well as many other observations for structure assemblies rely on so-called local determinism in the assembly \citep{Soloveichek-Winfree-2007}; in other words, for every two producible assemblies there is a larger assembly that contains both as sub-assemblies.
This property is sometimes called `directed assembly' or `determinism of the system' \citep{Patitz-survey}. 
In order to avoid the ambiguity and being guided by similar notions in other systems, here this property is also called {\it confluence}.

The standing question about the computational power of systems with non-cooperative binding (temperature 1 systems) has initiated several studies of these systems.
It has been observed that even small modifications of the model can provide universal computational power. 
For example, by considering three-dimensional tile assemblies that can add another layer of tiles (essentially having one array above the other) it was shown that confluent (directed)  non-cooperative assembly system has universal computational power \citep{cook+fu+schweller11soda}. 
It was also observed that by allowing one tile with a repelling glue 
(glue with strength $-1$), the confluent non-cooperative system becomes computationally universal \citep{negative-glue}. On the other side, if one allows certain tiles to be be added to the system in stages, then again, the system gains universal computing power \citep{stage-temp1-assembly}.
It was also observed that if  tiles are equipped with signals that activate glues stepwise, the system can simulate any temperature 2 system \citep{Daria2015}, that is, including the one that provides intrinsic universality of temperature~2 systems  \citep{IntrUniver}.

Infinite ribbon construction or snake tilings in non-cooperative (temperature 1)  binding systems were also given attention \citep{adleman+kari+kari+reishus02focs,adleman+kari+kari+reishus+sosik09siam,brijder+hoogeboom09tcs,kari02dlt}. It was observed that non-determinism, or non-confluence, also adds power to the system.
First it was shown that it is undecidable whether one can obtain an infinite ribbon (snake tiling) with a given non-deterministic system \citep{adleman+kari+kari+reishus+sosik09siam}.
This was achieved with simulating special type of Wang tiles and a space filling curve.
In this case, the notion of 
a `directed' system implies that the design of the tiles is accompanied with arrows that guide the direction of the assembly rather than the system being deterministic (or confluent). 
On the other side, one can use the snake tiles, and the space filling curve, to obtain a non-confluent system that can generate recognizable picture languages \citep{brijder+hoogeboom09tcs}.
And because recognizable picture languages contain the rectangular shapes that can be obtained from Wang tiles, which are known to have universal computational power \citep{Wang1975}, we have that non-determinism, i.e., non-confluence of the system together with a pre-defined condition on acceptable assembly provide universal computing power.

The limitation of a confluent (deterministic) non-cooperative (temperature 1) binding was first observed through so-called `pumpable' paths \citep{doty+patitz+summers11tcs}.
An infinite path is pumpable if there is a segment of the path that can extend into ultimately periodic within the assembled structure. 
By assuming that a system can have every sufficiently long  path `pumpable', it was observed that only a limited number of structures can be constructed with this system. 
In particular, in this case the finite assembly is either a `grid' or a `finite set of combs'.
This implies that the maximal assembly covers a semi-linear subset of the integer lattice.
It was also proved that confluent non-cooperative binding cannot simulate a trace of bounded Turing machine computation whose halting appears on the boundary of the computation \citep{meunier+woods17stoc}.
The paper also shows that such a system cannot be intrinsically universal, that is, there is no temperature-1 confluent aTAM that can simulate any other such a system.

A pumping lemma for temperature~1 confluent aTAM appears in \cite{meunier+regnault+woods20archive}, a sketch of the proof was accepted to STOC 2020 \cite{meunier+regnault+woods20stoc}.
The authors prove that if an assembly path starting from the seed is long enough, then there is a \emph{shield} structure beyond which there is a path that can be pumped and assembled.
This observation covers a large part of the long assembly path cases, but not all.
For example the `teeth' of the combs start away from the seed. 
The authors provide a bound for the size of the path to ensure the existence of a shield structure and thus obtain a pumpable path.
This assembly either exists in the maximal assembly or is `fragile' (for a non-deterministic system, the growth can be blocked by a previously assembled path).
In confluent system, no path can be fragile, so that long enough paths starting at the seed are pumpable, hence the final assembly has  ultimately periodic paths.
In a new paper \citep{meunier+regnault21} it was proven that the final structure  of a confluent temperature-1  is decidable.

This paper supplements and reinforces the result in \cite{meunier+regnault+woods20stoc,meunier+regnault+woods20archive} by providing a necessary and sufficient condition for a confluent (directed) non-cooperative tile assembly system to have an ultimately periodic assembly path. This is also an alternative proof of the result in~\cite{meunier+regnault+woods20stoc}. 
We show that such a path exists in an infinite maximal assembly if and only if there is an infinite assembly path that does not intersect an ultimately periodic path in the two-dimensional grid of $\IntegerSet^2$ (\RefLem{lem:nice}).
With this observation we prove that a maximal infinite assembly must contain an  ultimately periodic assembly path in the confluent aTAM at temperature 1.
Hence this is another proof of the main result in \citep{meunier+regnault+woods20stoc} using 
different tools such as  `free-paths' (\RefSec{path}) and superposition of free-paths called  `co-grow'~(\RefSec{sec:co-grow}) that are interesting on their own.

For the existence of ultimately periodic assembly paths we consider finite portions of paths, which we call \emph{off-the-wall paths}, that are bounded by a line in $\IntegerSet^2$.
These paths are used to show that the necessary and sufficient condition for existence of an ultimately periodic assembly path always holds (\RefTh{th:nice}).
Two main notions are used in the proofs: 
(a) left and right regions in the plane separated by a bi-infinite path and (b) superposition of two paths, which we call {\it co-grow} to obtain a new path that takes the `right-most' way of the two, similarly to the `right-priority' path in \citep{meunier+regnault+woods20stoc,meunier+woods17stoc}, except in our case co-grow is used as a function that produces a new `right-most' path of two that are in the input. 
The co-grow of two paths is possible if they are in a well defined region that has `no obstructions' which we call {\it non-causal}. Because the system is confluent, the intersection of the two paths during co-grow is always at a vertex that can be associated with only one tile type.


\section{Definitions}
\label{sec:definitions}

The set of integers from $a$ to $b$ is denoted \IntegerInterval{a}{b} ($a$ and $b$ can be infinite).
The two dimensional integer lattice $\IntegerSet^2$ is considered as a two dimensional grid, a periodic graph whose vertices are the 
elements of $\IntegerSet^2$ and two vertices $x$ and $y$ are connected by an edge if $||x-y||=1$.
A \emph{path} in $\IntegerSet^2$ is a simple path without repetition of the vertices and edges, it can be finite or (bi-)infinite.
A \emph{cycle} is a simple path whose first and last vertex are the same. 
The set of vertices visited by a path \Path is the domain of \Path and is denoted $\Domain{\Path}$.
We denote with $\Path_i$ the $i$th vertex visited by $\Path$ and for $a\le b$ we denote with $\Path_{\IntegerInterval{a}{b}}$ the segment $\Path_a\Path_{a+1}\cdots\Path_{b}$ of a path \Path.
We allow for one, or both $a$ and $b$ to be infinite.
The origin of ${\IntegerSet^2}$ is \ZZOrigin ($=(0,0)$).
The intersection of two paths is the set of vertices visited by both paths.

The set of unit vectors $\DirectionSet=\left\{
  \East{=}(1,0),
  \North{=}(0,1),
  \South{=}(0,-1),
  \West{=}(-1,0)
\right\}$ is called the set of \emph{directions}.
The vectors \East, \North, \South, and \West correspond to the east, north, south and west directions, respectively.

A graph $H$ that is a sub-graph of a graph $G$ is denoted by $H\SubGraphOf G$. 
We consider paths as graphs and the same notation is used for paths and subpaths.
Let $G$, $H$ and $H'$ be graphs such that $H\SubGraphOf G$ and $H'\SubGraphOf G$, $H\SubGraphCup H'$ denotes the subgraph that is union of $H$ and $H'$ in $G$.

The set of finite (resp. forward infinite, backward infinite, bi-infinite) sequences of elements, {\it words}, over alphabet $\DirectionSet=\{\West,\East,\South,\North\}$ is denoted $\DirectionSet^{*}$ (resp. $\DirectionSet^{\omega}$, $^{\omega}\DirectionSet$, $^{\omega}\DirectionSet^{\omega}$).
The union of $\DirectionSet^{*}$, $\DirectionSet^{\omega}$, $^{\omega}\DirectionSet$ and $^{\omega}\DirectionSet^{\omega}$ is denoted $\DirectionSet^{\IntegerSet}$.
The empty sequence is \EmptyWord.
We consider $d\in \DirectionSet$ as a symbol in the alphabet $\DirectionSet$ and a unit vector in $\IntegerSet^2$.

\subsection{Free Paths and Paths in $\IntegerSet^2$}\label{path}

We say that two paths $\Path$ and $\Path'$ in $\IntegerSet^2$ are \emph{equivalent} if $\Path'$ is a translation of $\Path$ in $\IntegerSet^2$.
The equivalence class of $\Path$, denoted \FreePathPath, is called a \emph{free path} associated with $\Path$.
The equivalence class is uniquely determined by a sequence of unit vectors $\kappa$, a word over $\DirectionSet$, (i.e., an element of $\DirectionSet^{\IntegerSet}$) such that $\kappa_i=d$ if and only if $\Path_{i+1}=\Path_{i}+d$.
We intermittently use the notion `free path' and a word notation $\kappa$ to represent both \FreePath{\Path} and a word in $\DirectionSet^{\IntegerSet}$.
The null free path is $\EmptyWord$ and \FreePath{\epsilon} is the set of vertices in $\IntegerSet^2$.
If $m_1$ is not forward infinite and $m_2$ is not backward infinite, then $m_1m_2$ designates their concatenation and represents a free path only if it corresponds to a path.
If $m$ is a finite free path, then $m^{\omega}=m m m\cdots$ is its infinite (forward) repetition, $^{\omega}m=\cdots m m m$ is its infinite backward repetition and $^{\omega}m^{\omega}=\cdots m m m\cdots$ is its bi-infinite repetition.
The set of cyclic rotations of a finite free path $m=d_1\cdots d_k$ is $\Rotate{m}=\{d_{i}d_{i+2}\cdots d_k d_1 d_2\cdots d_{i-1}|1\leq i \leq k\}$.

For any finite free path $m=d_1\cdots d_k\in\DirectionSet^*$, the associated \emph{displacement vector}, \Vect{m} (of $\IntegerSet^2$) is defined as the sum of its elements
$\Vect{m}=d_1+\cdots + d_k$.
Two finite free paths are \emph{collinear} if their associated displacement vectors are collinear.
There are infinitely many free paths associated to a given displacement vector of $\IntegerSet^2$ and they are all mutually collinear.

For a path \Path in $\IntegerSet^2$ and $A\in \Domain{\Path}$ we also use a notation $\Path=b.A.f$ where $b$ and $f$ are free paths such that $bf=\FreePath{\Path}$. We say that the free path $bf$ is {\it grounded}
at vertex $A$ and that $\Path$ is the resulting path.
That is, $b.A.f$ is an instance of a free path $bf$ such that the end vertex of the sub-path corresponding to $b$ is $A$, which is the first vertex to the subpath corresponding to $f$.
If any of these free paths, $b$ or $f$, is null, the notation simplifies to $b.A$, $A.f$ or just $A$.
Extending this notation, a path can also be denoted as a sequence of paths and free-paths; i.e., 
$q=f_1.\Path_1. f_2.\Path_2.\cdots$ where $f_i$'s are free paths and $\Path_i$'s are paths. In this case $q$ is the unique path instance in the equivalence class $f_1\FreePath{\Path_1}f_2\FreePath{\Path_2}\cdots$ `grounded' by the vertices in the domains of $\Path_1,\Path_2,\cdots$ (if it exists as a path).

If $A$ is a vertex of ${\IntegerSet^2}$ and $\Vect{v}$ is a vector in $\IntegerSet^2$, then $A+\Vect{v}$ is the vertex $A$ translated by \Vect{v}.
For any path, $bm.A=b.(A-\Vect{m}).{m}$ where $b\in D^*\cup {^{\omega}{D}}$ and $m\in D^*$. 
The \emph{reverse} of a free path $m=d_1d_2\cdots d_k$ is $\Reverse{m}=\Backward{d_k}\,\Backward{d_{k-1}}\cdots \Backward{d_1}$ where
$\Backward{\East}=\West$,
$\Backward{\North}=\South$, 
$\Backward{\South}=\North$, and
$\Backward{\West}=\East$.
In particular \Reverse{m} traverses the {free} path $m$ in reverse.

A free path $\alpha=mp^{\omega}$ for $m,p\in \DirectionSet^*$ and $p\neq\EmptyWord$ is called an \emph{ultimately periodic} free path, the prefix $m$ is called the \emph{transient part} of $\alpha$ and $p^{\omega}$ is the \emph{periodic part} of $\alpha$.
Similarly, the path $\Path=A.mp^{\omega}$ is an \emph{ultimately periodic} path, $A.m$ is the \emph{transient part} of $\Path$ and $(A+\Vect{m}).p^{\omega}$ is the \emph{periodic part} of $\Path$.
A \emph{periodic} (free) path is a (free) path whose transient path is $\epsilon$.

Let \Path be a path and $\Path_i$ be a legal vertex of \Path.
The notation \PathFrom{\Path}{i} (resp. \PathAfter{\Path}{i}) corresponds to the subpath of \Path starting at vertex $\Path_i$ till the end of $\Path$, with $\Path_i$ included (resp. excluded).
The notation \PathTo{\Path}{i} (resp. \PathBefore{\Path}{i}) corresponds to the subpath of $\Path$ up to vertex $\Path_i$, with $\Path_i$ included (resp. excluded).
For example, for a bi-infinite path \Path,
$\PathFrom{\Path}{i}=\Path_{\IntegerInterval{i}{+\infty}}$,
$\PathAfter{\Path}{i}=\Path_{\IntegerInterval{i+1}{+\infty}}$,
$\PathTo{\Path}{i}=\Path_{\IntegerInterval{-\infty}{i}}$, and
$\PathBefore{\Path}{i}=\Path_{\IntegerInterval{-\infty}{i-1}}$.

\subsection{Regions}

A \emph{region}, $R$, is a connected subgraph of $\IntegerSet^2$.
For a vertex $v\in \IntegerSet^2$, the \emph{neighborhood of $v$}, $\mathcal{N}(v)$, is the subgraph of $\IntegerSet^2$ induced by the nine vertices at $x$-distance and $y$-distance at most $1$ from $v$.
A \emph{boundary} vertex for a region $R$ is a vertex $A\in R$ such that $\mathcal{N}(A) \not\SubGraphOf R$.
The \emph{boundary of $R$}, denoted $\Boundary R$, is the subgraph of $R$ induced by the sets of its boundary vertices.
The interior of $R$, denoted $\mathring{R}$, is the complement (with complement taken out of vertices and edges) of $\Boundary R$ in $R$, that is $\mathring{R}=R\setminus\Boundary{R}$.

Let $\Path$ be a cycle or a bi-infinite path in $\IntegerSet^2$.
Then, by Jordan curve theorem, $\Path$ defines two regions in the plane whose intersection is $\Path$ itself.
We distinguish these two regions as `left' and `right' as described below.

Since all paths are in dimension $2$, a path $\Path$ can be considered oriented by orienting the edges from $\Path_i$ to $\Path_{i+1}$ so that its left side can be defined.
For $d,d'\in \DirectionSet$ we say that $d'$ is {\it to the left of} $d$ or $d$ is {\it to the right of} $d'$ if $(d,d')\in\{(\North,\West),(\West,\South),(\South,\East),(\East,\North)\}$.
A vertex $A$ in $\IntegerSet^2$ is \emph{directly to the left of $\Path$} if it does not belong to \Path and there are $i\in\NaturalSet$ and a direction $d'$ to the left of $d$ such that $\Path_i+d=\Path_{i+1}$ and $\Path_i+d'=A$, or $\Path_{i-1}+d=\Path_i$ and $\Path_i+d'=A$.
In this case the edge $(A,\Path_i)$ is \emph{directly to the left} of $\Path$.
Similarly we define a vertex, and an edge \emph{directly to the right of $\Path$}.

The \emph{left region} of a path $\Path$ is the subgraph of $\IntegerSet^2$ consisting of vertices $A$ that are either in \Domain{\Path}, or there is a path $A.m$ for some free path $m$ that ends at $\Path$ with an edge directly to the left of $\Path$ and does not intersect with $\Path$ in any other vertex.
The \emph{right region} of $\Path$ is defined similarly.
Because $\Path$ is bi-infinite or a cycle, the left and right regions are well defined, and their intersection is $\Path$.

A path $\Path$ is \emph{inside a region} $R$ if $\Path\SubGraphOf R$.
It is \emph{strictly inside} $R$ if $\Path\SubGraphOf \mathring{R}$.

Let $\Vect{v}=(p,q)$ ($p,q\in\IntegerSet$) be a vector and consider the line $\ell$ defined with $\ZZOrigin+s\Vect{v}$ for $s\in \mathbb{R}$ in the Euclidean plane $\mathbb{R}^2$.
Then, we also define the right and the left regions of 
$\ell$ such that $A=(a,b)\in \IntegerSet^2$ is in the {\it left region of $\ell$} if the dot product $A\cdot\Vect{v}_\perp$ (where $\Vect{v}_\perp=(-q,p)$) is positive or null.
Similarly, $A$ is in the {\it right region of $\ell$} if $A\cdot\Vect{v}_\perp≤0$. For an arbitrary line $\ell'$ that is a translation of $\ell$ defined with $B+s\Vect{v}$ 
for $B\in \IntegerSet^2$, $s\in \mathbb R$, the vertex $A\in \IntegerSet^2$ is in the {\it right} region of $\ell'$ if $A-B$ is in the right region of $\ell$. The {\it left} region of $\ell'$ is defined analogously. In particular, the points of $\IntegerSet^2$ that belong to $\ell$ (or $\ell'$) belong to both, the left and the right region of $\ell$ (or $\ell'$). Observe that if $R$ is the right region of $\ell$ (or $\ell'$), then $\Boundary R$ is a bi-infinite periodic path because $\Vect{v}$ has integer coordinates. Similarly the boundary of the left region is a bi-infinite periodic path (generally different from the boundary of the right region). 

Let $A\in \IntegerSet^2$ be in the left region of $B+s\Vect{v}$. The {\it ribbon} between points $A$ and $B$ in $\IntegerSet ^2$, $A\not = B$ in direction $\Vect{v}$ is the intersection of the left region of $B+s\Vect{v}$ and the right region of $A+s\Vect{v}$ and is denoted $(A,B, \Vect{v})$. 
Directly from the definition it follows that if $C$ is a vertex in the ribbon $(A,B,\Vect{v})$ then all points of the integer lattice $\IntegerSet^2$ that lie on the line $C+s\Vect{v}$ are also in the ribbon. 

\begin{lemma}[double implies periodic]
  \label{lem:double-pumpable}
  Let $m$ be a non null finite free path.
  If $m^{2}=mm$ is a free path then so is $^{\omega}m^{\omega}$.
\end{lemma}

\begin{proof}
  Let $m=d_1d_2\cdots d_t$ be a free path such that $m^{2}$ is a free path.
  We denote $m_1=0$ and $\Vect{m_r}=\Vect{d_1}+\cdots+\Vect{d_{r-1}}$.
  Because $m$ is non null and is a free path, it follows that $\Vect{m}\not=\Vect{0}$.
  Suppose that $k$ is such that $\ZZOrigin.m^{k}$ is a path, but $\ZZOrigin.m^{k+1}$ is not a path but a walk.

  Let $A$ and $B$ be such that all vertices of the walk $\ZZOrigin.m^{k+1}$ are included in the ribbon $(A,B,\Vect{m})$ and $||A-B||$ is minimal. If $A=B$, because the ribbon contains the path $\ZZOrigin.m^{k}$ (connected subgraph), then $\Vect{m}$ is either horizontal, or vertical, in which case $\ZZOrigin.m^{k+1} = d\cdots d$ for some direction $d\in \DirectionSet$, and the lemma holds.
  Hence we assume $0<||A-B||$. 
  Because by definition all vertices of the walk $\ZZOrigin.m^{k+1}$ are included in the ribbon, and by the minimality condition, the vertices of $\ZZOrigin.m^{k+1}$ are either boundary vertices for the ribbon or they are strictly inside the region defined by the ribbon.

  By the minimality of $||A-B||>0$, there are integers $i$ and $j$ with $\Vect{m_i}=\Vect{d_1}+\cdots+\Vect{d_{i-1}}$ and a sub-path $\Vect{m_i}.m_{\IntegerInterval{i}{j}}$ of $\ZZOrigin.m$ that splits the ribbon in two parts, that is, $\Vect{m_i}.d_i$ is at one boundary of the ribbon (say the boundary of the right region of $A+s\Vect{m}$), and $\Vect{m_j}.d_j$ is on the other boundary of the ribbon (the boundary of the left region of $B+s\Vect{m}$). 
  
  If $i<j$, we can choose $i$ and $j$ such that none of the vertices $\Vect{m_r}.d_r$ are on the lines $A+s\Vect{m}$ and $B+s\Vect{m}$ for $i<r<j$.
  Any bi-infinite path within the ribbon passing from one part of the ribbon to the other must intersect with a vertex from $\Vect{m_i}.m_{\IntegerInterval{i}{j}}$. 
  Moreover, for every integer $l$, the vertices of $(l\Vect{m}).m$ are included in the ribbon and $(l\Vect{m}+\Vect{m_i}).m_{\IntegerInterval{i}{j}}$ splits the ribbon in two parts.

  Being $\ZZOrigin.m^k$ a path, but $\ZZOrigin.m^{k+1}$ a walk and not a path, $(k\Vect{m}).m$ must intersect $\ZZOrigin.m^k$.
  This means that it either intersects the tail of $(k{-}1)\Vect{m}.m$ or it must cross the sub-path $((k{-}1)\Vect{m}+\Vect{m_i}).m_{\IntegerInterval{i}{j}}$.
  Hence it has to intersect $(k{-}1)\Vect{m}.m$, implying that $(k{-}1)\Vect{m}.m^2$ is not a path, but a walk. This is in contradiction with our assumption that $m^2$ is a free path.
\end{proof}

\subsection{Co-grow}
\label{sec:co-grow}

We describe a method of superimposing two free paths to form a new free path that is in some sense the `rightmost' portion of both. The idea is similar to taking the ``right-priority'' in \citep{meunier+woods17stoc}.
Here we define a general method to take two bi-infinite free paths that intersect and obtain a forward finite, or infinite free path starting at one of the intersection points that coincides with at least one of the paths and lies within the right regions of both paths. We call this combined new path as `co-grow' of both. 
The co-grow is a free path and becomes a path once we ground it at a starting vertex. 

\begin{definition}[co-grow]\label{co-grow}
  Let $bf$, $b'f'$ be free bi-infinite paths (where $b, b'$ are backward infinite and $f, f'$ are forward infinite)
  such that 
  $f$ and $f'$ start with the same direction.
  Let $R$ (resp. $R'$) be the right region of $b.\ZZOrigin.f$ (resp. $b'.\ZZOrigin.f'$).
  The \emph{(right) co-grow of the free paths $b$, $f$, $b'$ and $f'$}, denoted $\hat f=\CoGrow{b}{f}{b'}{f'}$ is a forward, possibly infinite, maximal free path $\hat f$ corresponding to the path $\ZZOrigin.\hat f$ defined
  inductively as follows:
  \begin{itemize}
  \item 
    $\ZZOrigin.\hat f_1=\ZZOrigin.d$ 
    where the 
    direction $d=f_1=f'_1$ is the common initial direction of $f$ and $f'$;
  \item if $\ZZOrigin.\hat f_1\cdots \hat f_i$ is defined and it is a path from 
    $\ZZOrigin$ to $A$, then $\hat f_{i+1} =d$ is defined if 
    \begin{itemize}
    \item $d$ is the rightmost direction with respect to $\hat f_i$ such that $A.d\SubGraphOf\ZZOrigin.f$ or 
      $A.d\SubGraphOf\ZZOrigin.f'$; 
    \item $\ZZOrigin.\hat f_1\cdots\hat f_{i+1}=\ZZOrigin.\hat f_1 \cdots \hat f_i d$ is a path that is a subgraph of 
      $R\sqcap R'$.
    \end{itemize}
  \end{itemize}
\end{definition}

\begin{figure}[hbt]
  \centering
  \subcaptionbox{\label{fig:cog:regions:a}%
    Simple co-grow%
  }{%
    \includegraphics[scale=.9]{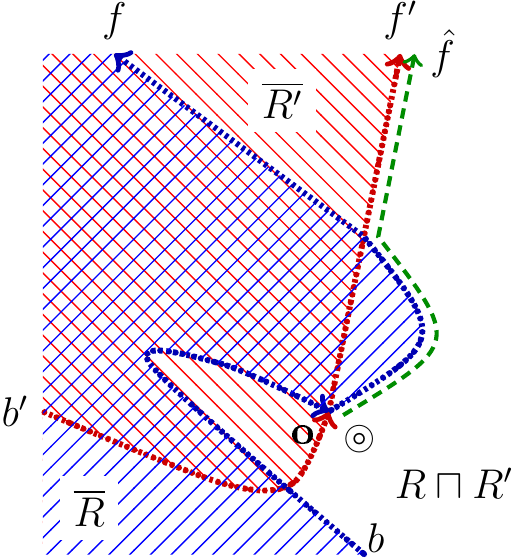}%
  }%
  \hskip.5cm%
  \subcaptionbox{%
    \label{fig:cog:regions:b}%
    finite co-grow%
  }{%
    \includegraphics[scale=.9]{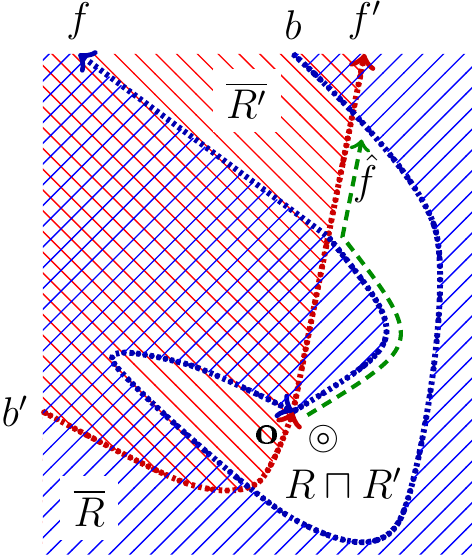}%
  }%
  \hskip.5cm%
  \subcaptionbox{%
    \label{fig:cog:regions:c}%
    unvisited bubble%
  }{%
    \includegraphics[scale=.9]{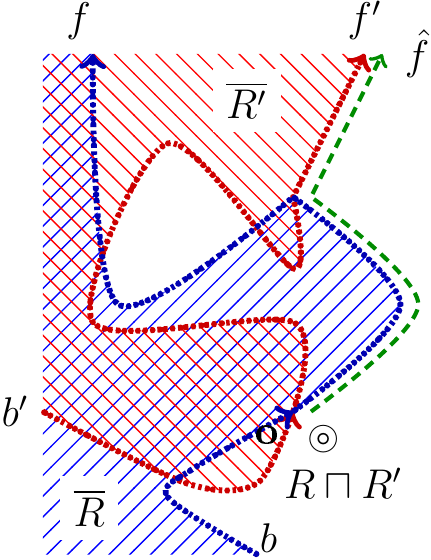}%
  }
  \caption{co-grow. (a) The dashed (green) line indicates the path $\hat f= \CoGrow{b}{f}{b'}{f'}$ that is a co-grow of $f$ (in blue) and $f'$ (in red). 
    Examples where the intersection of the right region of the red path and the blue path is (b) finite and (c) not connected.
    However, the co-grow of the two paths (in green) bounds the connected component of the region intersection that contains the starting vertex.
    In case of (b) $\hat f$ is finite, not necessarily bounding the whole component.
    It cannot be extended because there is no rightmost direction that belongs to $f$ or $f'$ that is also in $R\SubGraphCap R'$.
    In case of (c) the region bounded by $\hat f$ is potentially infinite, and so may be $\hat f$.}
  \label{fig:cog:regions}%
\end{figure}

The notation of paths and regions in co-grow remain fixed for the rest of the current section.
Paths, regions and expected free path $\hat f$ are illustrated in \RefFig{fig:cog:regions}.
The symbol \CoGrowStart is used to indicate that start of the co-grow.
The free paths $b$ and $b'$ do not take part in the construction of $\hat f$ but are used to define the two right regions $R$ and $R'$ producing the boundaries $b.\ZZOrigin.f$ and $b'.\ZZOrigin.f'$ that limit the
extension of the co-grow (see \RefFig{fig:cog:regions:a}).
If paths, instead of free paths, are used as arguments for co-grow, then the co-grow is
assumed to be taken with the associated free paths.
The left co-grow of $b$, $f$, $b'$ and $f'$, denoted $\LCoGrow{b}{f}{b'}{f'}$ is defined in a similar way by considering the leftmost directions and the left regions of these paths. In the rest of our exposition we use simply `co-grow' for the right co-grow.

The `growing direction' of the co-grow at any vertex coincides with the direction of at least one of the paths $\ZZOrigin.f$ and $\ZZOrigin.f'$ at the same location, and if they intersect at that vertex it always takes the `rightmost' path of the two paths at the current vertex.
A way to obtain the path $\hat f=\CoGrow{b}{f}{b'}{f'}$ is to start with $\epsilon$ from the origin and follow the direction of both paths $f$ and $f'$ until one of the paths takes a direction different from the other.
At that point one follows the path that takes the rightmost direction, until the point when both paths intersect.
In a sense, in between any two intersections of paths $\ZZOrigin.f$ and $\ZZOrigin.f'$ one follows the path that is strictly to the right of the other.

\begin{figure}[hbt]
  \centering%
  \small%
  \subcaptionbox{\label{fig:cog:regions:isthmus f f'}%
    isthmus between $f$ and $f'$ that is avoided by $\hat f$
  }{%
    \includegraphics[scale=1]{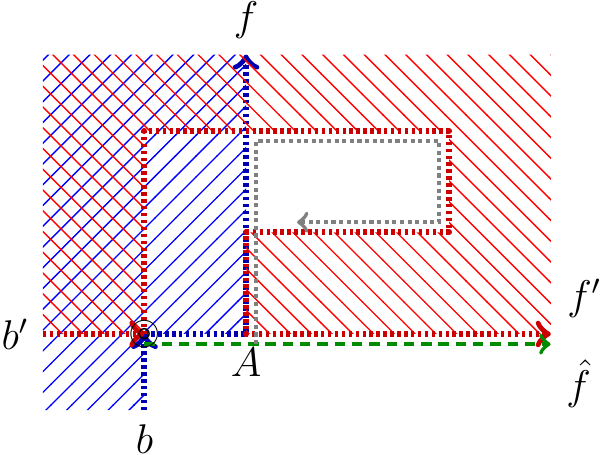}%
  }
  \hskip 1cm
  \subcaptionbox{\label{fig:cog:regions:isthmus f b'}%
    isthmus between $f$ and $b'$ that is part of $\hat f$%
  }{%
    \includegraphics[scale=1]{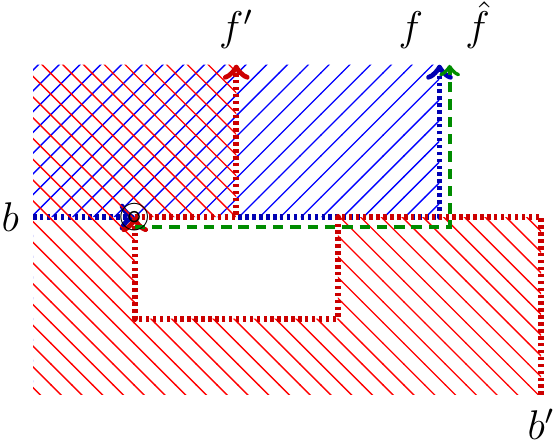}%
  }
  \caption{Isthmus and the construction of $\hat f$.}
  \label{fig:cog:isthmus}%
\end{figure}

The co-grow produces a `maximal' free path $\hat f$ in the boundary of the region $R\SubGraphCap R'$ that contains the origin.
Observe that $R\SubGraphCap R'$ may be finite or not be connected (see figures\,\ref{fig:cog:regions:b} and \ref{fig:cog:regions:c}), but the path $\hat f$ (in green) that is a co-grow of two paths (red and blue), lies on the boundary of the connected component that contains the starting vertex.
In the case of \RefFig{fig:cog:regions:b} the region that is bounded by the co-grow is finite and $\hat f$ is a finite path (not bounding the region completely).
In the case of \RefFig{fig:cog:regions:c}, the co-grow is possibly infinite. 
A component in $R\SubGraphCap R'$ could be connected only through a path (isthmus in graph theory).
The rightmost extension in the definition is important as illustrated in \RefFig{fig:cog:regions:isthmus f f'} where the condition prevents $\hat f$ from taking an isthmus between $f$ and $f'$ at point $A$ and get trapped in a finite region (dotted path).
Note that there may be an isthmus between $f$ and $b'$ as shown in \RefFig{fig:cog:regions:isthmus f b'}, but this condition may not necessarily enforce finiteness of $\hat f$ because $b'$ is part of the right region $R'$, as well as $f$ is a part of $R$ (in comparison with~\RefFig{fig:cog:regions:b} where $f$ crosses into the complement of $R'$).
The situation when the co-grow produces an infinite free path is of our interest; therefore we have the following lemma.

\begin{lemma}
  \label{inf-co-grow}%
  Let $bf$, $b'f'$ be bi-infinite free paths such that $f$ and $f'$ start in the same direction, and let $\hat f=\CoGrow{b}{f}{b'}{f'}$.
  Let $R$ (resp. $R'$) be the right region of $b.\ZZOrigin.f$, (resp. $b'.\ZZOrigin.f'$).
  If there is a forward infinite path in $R\SubGraphCap R'$ starting at \ZZOrigin, then $\hat f$ is infinite.%
\end{lemma}

\begin{proof}
  Let $K$ be the connected component of $R\SubGraphCap R'$ that contains $\ZZOrigin$.
  Assume that there is a forward infinite path in $R\SubGraphCap R'$ starting at \ZZOrigin.
  The connected component $K$ then must be infinite.
  
  We say that a \emph{path $q$ crosses into some region $S$ over some path $q'$} when there are two indices $i\le j$ such that: 
  $q_{\IntegerInterval{i}{j}}\SubGraphOf q'$
  and one of $q_{i-1}$ and $q_{j+1}$ belongs to the interior of $S$, while the other does not belong to $S$.
  If $q_i$ intersects $q'$ and one of $q_{i+1}$ or $q_{i-1}$ is in the interior of $S$, then $q_i$ is called {\it the crossing point into $S$ over $q'$}.
  
  Since $K$ contains the origin and is infinite, $b'.\ZZOrigin$ cannot cross into $K$ over $\ZZOrigin.f$.
  Otherwise there is a cycle between the crossing point into $K$, following $b'$ to \ZZOrigin, then following $f$ ending at the crossing point into $K$.
  This cycle contains the origin and would bound $K$ inside a finite region, but since $K$ is infinite, this is a contradiction.
  Similarly, $b.\ZZOrigin$ cannot cross into $K$ over $\ZZOrigin.f'$.

  We denote $\ZZOrigin.f$ by $\Path$, $\ZZOrigin.f'$ by $\Path'$ and $\ZZOrigin.\hat{f}$ by $\hat{\Path}$ where $\hat{f}$ is the co-grow of $b$, $f$, $b'$, and $f'$.
  
  Because every vertex in $\hat{\Path}$ belongs to at least one of $\Path$ or $\Path'$, let $\{j_n\}_n$ and $\{i_n\}_n$ be the respective indexes of vertices of $\Path$ and $\Path'$ intersecting $\hat{\Path}$ listed in the order they are encountered.
  We prove that at least one of these sequences is infinite by showing that they must be increasing sequences and the rightmost direction of co-grow always exists. 
  By Def.~\ref{co-grow}, $\hat{f}$ has at least one edge because $f$ and $f'$ start with the same direction and so $j_0=i_0=0$ and $j_1=i_1=1$. 
  
  Inductively, for $k\ge 1$, suppose $\ZZOrigin.\hat{f}_1 \cdots \hat{f}_k=\hat{\Path}_{\IntegerInterval{0}{k}}$ is defined intersecting $\Path$ at vertices with indexes $j_0,j_1,\ldots,j_s$ (on \Path), and intersecting $\Path'$ at indexes $i_0,i_1, \ldots, i_{t}$ (on $\Path'$).
  Assume further that if $\hat{\Path}_m=\Path_{j_n}$ and $\hat{\Path}_{m'}=\Path_{j_{n+1}}$, then $m<m'$ implies $j_n<j_{n+1}$, i.e., $\hat{\Path}_{\IntegerInterval{0}{k}}$ meets vertices of $\Path$ with increasing indexes, that is, $\hat{f}$ always follows the same direction as $f$ when using edges from $f$.
  Suppose the same holds for $\Path'$ and $f'$. 
  
  Because $f$ and $f'$ are forward infinite, at least one of the edges of $f$ or $f'$ is rightmost starting at $\hat{\Path}_k$, and the co-grow $\hat{\Path}$ could extend with that edge.
  The only way that co-grow's $(k+1)$-st edge would not exist is when the rightmost edge exits the region $R\SubGraphCap R'$ (the second condition of the inductive step in Def.~\ref{co-grow}).
  This happens when taking the next rightmost direction crosses one of $b.\ZZOrigin$ or $b'.\ZZOrigin$, which, as observed above, cannot happen.
  
  Assume that rightmost direction from $\hat{\Path}_k$ is a direction using $f$ (the case when the co-grow uses $f'$ is similar) with $\hat{\Path}_k=\Path_{j_s}$ and $\hat{\Path}_{k+1}=\Path_{j_{s+1}}$.
  The last direction of $\hat{\Path}$ joining $\hat{\Path}_{k-1}$ and $\hat{\Path}_k$ either follows $f$ or not.
  Case 1: if $\hat{\Path}_{k-1}=\Path_{j_{s-1}}$, and $\hat{\Path}_{k}=\Path_{j_s}$, that is, the last direction reaching $\hat{\Path}_k$ taken by $\hat{\Path}$ is following $f$, then $\Path_{j_{s-1}}\Path_{j_s}\Path_{j_{s+1}}$ is a two edge sub-path of $\Path$. 
  By inductive hypothesis $j_{s-1}<j_s$ implies that $j_{s-1}+1=j_s$, so it must be $j_s+1=j_{s+1}$, i.e., $j_s<j_{s+1}$.
  
  \begin{figure}[hbt]
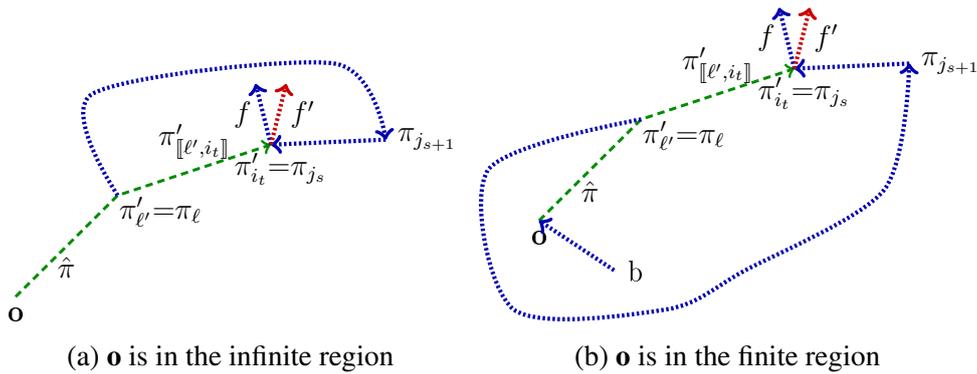

    \centering\small
    \newcommand{\Cas}[1]{
      \subcaptionbox{\label{fig:lem3:impossible:orientation:#1}%
        \ZZOrigin is in the #1 region}{%
        \includegraphics[scale=.93]{fig_cogrow_impossible_#1.pdf}
      }%
    }
    \Cas{infinite}
    \Cas{finite}
    \caption{Impossible orientation for $f$ (in blue) when the last direction taken by $\hat{\Path}$ coincides with $f'$ (in green).}
    \label{fig:lem3:impossible:orientation}
  \end{figure}

  Case 2: suppose $\hat{\Path}_{k-1}=\Path'_{i_{t}-1}$, and 
  $\hat{\Path}_{k}=\Path'_{i_{t}}$, that is, the last direction reaching $\hat{\Path}_k$ taken by $\hat \Path$ is following $f'$.
  Because $\hat{\Path}_k\hat{\Path}_{k+1}$ is an edge in $\Path$, we have $\Path'_{i_{t}}=\Path_{j_s}$, and either $
  j_s-1=j_{s+1}$ ($\hat f$ follows $f$ in the opposite direction) or $j_s+1=j_{s+1}$ ($\hat f$ follows $f$ in the same direction).
  We show that it must be the latter case. 
  By contradiction, assume $j_s-1=j_{s+1}$. 
  Let $\Path_\ell=\Path'_{\ell'}$ be vertex on $\hat{\Path}$ that is
  an intersection point of $\ZZOrigin.f$ and $\ZZOrigin.f'$ such that $\Path'_{\IntegerInterval{\ell'}{i_{t}}}\SubGraphOf\hat{\Path}$ has no other intersections between these two paths, except the endpoints.
  The region bounded by the cycle $\Path'_{\IntegerInterval{\ell'}{i_{t}}}.\Reverse{\FreePath{\Path_{\IntegerInterval{j_s}{\ell}}}}$ either contains $\ZZOrigin$ or not.
  The two possible cases are depicted in Figs.\,\ref{fig:lem3:impossible:orientation:infinite} and \ref{fig:lem3:impossible:orientation:finite}. 
  In the former case (Fig.\,\ref{fig:lem3:impossible:orientation:infinite}), the cycle $\Path'_{\IntegerInterval{\ell'}{i_{t}}}.\Reverse{\FreePath{\Path_{\IntegerInterval{j_s}{\ell}}}}$ would bound a finite region that must include the forward infinite \PathAfter{\Path}{j_s}.
  In the latter case (Fig.\,\ref{fig:lem3:impossible:orientation:finite}), the same cycle bounds a finite region containing \ZZOrigin, so the backward infinite path $b.\ZZOrigin$ must cross either $\ZZOrigin.f$ or $\ZZOrigin.f'$ which, as observed above, cannot happen. Therefore, it must be that $j_s+1=j_{s+1}$, and $\hat f$ follows the same direction as $f$.

  Because the co-grow extends at every step, and $\{j_n\}_n$ and $\{i_n\}_n$ are increasing sequences, at least one of the sequence of intersection indexes $\{j_n\}_n$, or $\{i_n\}_n$ is infinite, and hence, $\hat f$ is infinite.
\end{proof}



\subsection{Tile Assembly System}

Let \GlueSet be an a finite set called an \emph{alphabet} whose elements are symbols that we will also call \emph{glues}.
A \emph{tile type} is a map $t: \DirectionSet \to \GlueSet$.
We use the notation $t_d$ for the value (glue) of $t$ in direction $d$. 

A \emph{(temperature 1) tile assembly system} (TAS) is a pair $\TAS =(\TileSet,\Seed)$ where $\TileSet$ is a finite set of tile types and \Seed is a tile type (not necessarily in \TileSet) called the \emph{seed}.
In order to separate the seed from the tiles used in the assembly, we assume that $\Seed\not\in \TileSet$\footnote{The seed can also be taken to be an assembly larger than a singleton tile (e.g.~\citep{IntrUniver})}. 

For a TAS with set of tile types \TileSet and seed \Seed, an \emph{assembly} over \TAS is a partial map $\Assembly:\IntegerSet^2\to \TileSet\cup\{\Seed\}$ where $\alpha^{-1}(\Seed)$ is empty or a singleton.
The domain of \Assembly is the set of points of $\IntegerSet^2$ for which $\Assembly$ is defined, and is denoted $\Domain{\Assembly}$.
The \emph{binding graph} of $\Assembly$ is a subgraph of the lattice $\IntegerSet^2$ with vertices $\Domain{\Assembly}$ such that for $A,A'\in \Domain{\Assembly}$ there is an edge with endpoints $A$ and $A'$ if and only if $A+d=A'$ for some direction unit vector $d$ and 
$\Assembly(A)_d=\Assembly(A')_{-d}$.
The assembly $\Assembly$ is \emph{stable} if its binding graph is connected. 
An assembly $\Assembly$ is \emph{producible} in \TAS if it is stable and $\Assembly(0,0)$ is the seed.
The seed appears in $\alpha$ only at the origin.

Note that although neighboring vertices $A,A'\in\IntegerSet^2$ with $A+d=A'$ may be in the domain of $\alpha$ it may happen that $\alpha(A)_d\not=\alpha(A')_{-d}$.
In this case the tiles $\alpha(A)$ and $\alpha(A')$ mismatch in direction $d$, and the binding graph of $\alpha$ has no edge between $A$ and $A'$. 

A producible assembly $\Assembly$ over \TAS is said to be \emph{an assembly path} if its binding graph has as a subgraph a path $\Path$ with $\Path_0=(0,0)$ that visits all vertices of $\Assembly$'s binding graph, i.e., $\Domain{\Assembly}=\Domain{\Path}$. 
In other words $\Assembly$ is an assembly path if its binding graph has a spanning tree that is a path in $\IntegerSet^2$ starting at the origin. 
In this case we write $\Assembly=\Assembly_\Path$.
We note that for an assembly path a path $\Path$ such that $\Assembly=\Assembly_\Path$
may not be unique. For example, for an assembly path $\Assembly$ that has as a binding graph the unit square, we have two paths $\Path_1=\ZZOrigin.\East\North\West$ and $\Path_2=\ZZOrigin.\North\East\South$ such that $\Assembly=\Assembly_{\Path_1}=\Assembly_{\Path_2}$.
We extend the notions for ultimately periodic paths to ultimately periodic assembly paths.
We say that an assembly path $\Assembly$ is ultimately periodic if there is a path $\Path$ with $\Assembly=\Assembly_{\Path}$ and there are free paths $m$ and $p=p_1\cdots p_k$ such that for all $i\ge 0$, and all $s=1,\ldots,k$, $\Assembly(\Vect{m}+i\Vect{p}+\Vect{p_1\cdots p_s})=\Assembly(\Vect{m}+\Vect{p_1\cdots p_s})$.

We introduce the following property for tile assembly systems that we show holds for all confluent systems and helps to characterize the assemblies obtained in these systems.

\begin{definition}
  Two assemblies $\Assembly$ and $\Assembly'$ are \emph{compatible} if for all $v\in \Domain{\Assembly}\cap \Domain{\Assembly'}$ we have that $\Assembly(v)=\Assembly'(v)$. 
\end{definition}

\begin{definition}[confluent or directed]
  A tile assembly system \TAS is \emph{confluent} if every two producible assemblies $\Assembly$ and $\Assembly'$ are compatible.
\end{definition}

An assembly $\beta$ is \emph{maximal} for a system \TAS if for any other assembly $\alpha$ satisfying $\Domain{\beta}\subseteq \Domain{\alpha}$ we have that $\Domain{\alpha}=\Domain{\beta}$.
In a confluent system, any two assembly paths can `coexist' within a larger assembly because all intersections of their domains are mapped to the same tiles by both paths. 

\begin{lemma}\citep{doty+patitz+summers11tcs}
  If \TAS is confluent then there is a unique maximal producible stable assembly $\AlphaMax$ such that for every other stable assembly $\alpha$, $\Domain{\alpha}\subseteq\Domain{\AlphaMax}$.
\end{lemma}

\noindent
\emph{Notation.} In the rest of the paper we assume that \TAS is a fixed confluent tile assembly system that produces an infinite stable maximal assembly denoted \AlphaMax.
We introduce several straightforward lemmas that are used later in the text.

\begin{lemma}
  \label{path-assembly}%
  In a confluent system if for an assembly path $\Assembly$ there is a forward infinite ultimately periodic path \Path such that $\Assembly=\Assembly_{\Path}$, then $\Assembly$ is ultimately periodic.
\end{lemma}

\begin{proof}
  Suppose $\Assembly=\Assembly_{\Path}$ and $\Path=\ZZOrigin.mp^\omega$ is ultimately periodic. Observe that because \TAS is finite, there are $i<j$ such that $\Assembly_{\Path}(\Vec{m}+i\Vec{p})=\Assembly_{\Path}(\Vec{m}+j\Vec{p})$.
  Let $q$ be any prefix of $p^{j-i}$. Because of confluence of \TAS, it must be that $\Assembly_{\Path}(\Vec{m}+i\Vec{p}+\Vect{q})=\Assembly_{\Path}(\Vec{m}+j\Vec{p}+\Vect{q})$.
  Hence, $\Assembly$ is ultimately periodic since 
  for all $k\ge 0$, we have 
  $\Assembly(\Vect{m'}+k\Vect{p'}+\Vect{q})=\Assembly(\Vect{m'}+\Vect{q})$ where $m'=mp^i$ and $p'=p^{j-i}$.
\end{proof}

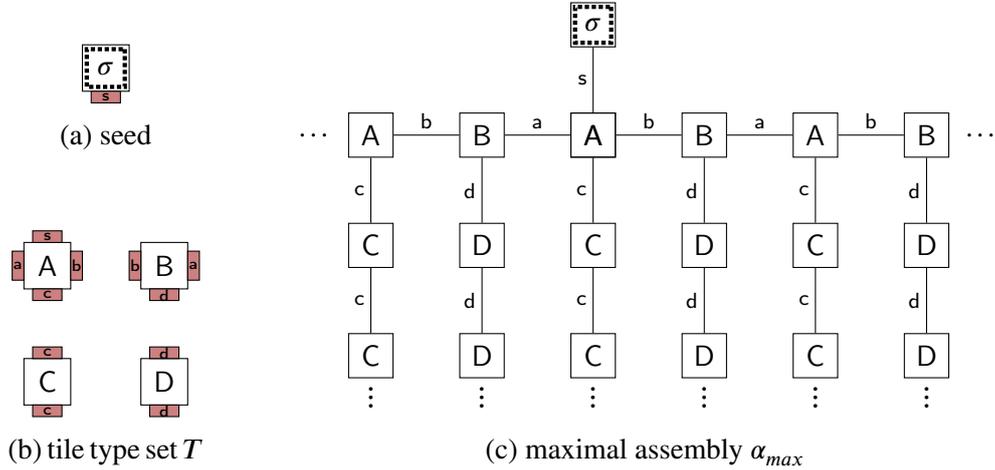
\begin{figure}[hbt]
  \centering%
  \small%
  \SetUnitlength{1.9em}%
  \begin{tabular}[b]{c}
    \SetUnitlength{2em}%
    \subcaptionbox{seed\label{fig:seed}}{%
    \qquad%
    \begin{tikzpicture}%
      \Tile[S]{0,4}{\Seed}%
      \GlueBot[s]{S}%
      \SEED{0,4}%
    \end{tikzpicture}%
    \qquad\mbox{}%
    }%
    \\[4em]
    \SetUnitlength{2em}%
    \subcaptionbox{tile type set \TileSet\label{fig:tas}}{%
    \begin{tikzpicture}
      \Tile[A]{0,2}{A}
      \GlueTop[s]{A}
      \GlueLe[a]{A}
      \GlueRi[b]{A}
      \GlueBot[c]{A}
      \Tile[B]{2,2}{B}
      \GlueLe[b]{B}
      \GlueRi[a]{B}
      \GlueBot[d]{B}
      \Tile[D]{2,0}{D}
      \GlueTop[d]{D}
      \GlueBot[d]{D}
      \Tile[C]{0,0}{C}
      \GlueTop[c]{C}
      \GlueBot[c]{C}
    \end{tikzpicture}
    }%
  \end{tabular}
  \qquad%
  \subcaptionbox{maximal assembly \AlphaMax\label{fig:assembly}}{%
    \begin{tikzpicture}%
      \Tile[S]{8,4}{\Seed}
      \SEED{8,4}
      \begin{scope}{opacity=0}
        \Tile[A]{8,2}{A}
      \end{scope}
      \LinkBotTop[s]{S}{A}
      \foreach \X in {1,2,3} {
        \begin{scope}[shift={(4*\X,0)}]
          \Tile[A]{0,2}{A}
          \ifnodedefinedcurrpic{B}{
            \LinkLeRi[a]{A}{B}}{}
          \Tile[B]{2,2}{B}
          \LinkLeRi[b]{B}{A}
          \Tile[C]{0,0}{C}
          \LinkBotTop[c]{A}{C}
          \Tile[D]{2,0}{D}
          \Tile[D1]{2,-2}{D}
          \LinkBotTop[d]{B}{D}
          \LinkBotTop[d]{D}{D1}
          \Tile[C1]{0,-2}{C}
          \LinkBotTop[c]{C}{C1}
          \draw (0,-2.75) node {\vdots} ;
          \draw (2,-2.75) node {\vdots} ;
        \end{scope}
      };
      \draw (3,2) node {\dots} ;
      \draw (15,2) node {\dots} ;
    \end{tikzpicture}
  }
  \caption{Example of \TAS, the east, west sides of tiles C and D as well as north of B have glues that do not match any other glues, hence are not indicated.}
  \label{fig:tas-example}
\end{figure}

\begin{example}
  \label{example:confl}%
  The tile set in figs. \ref{fig:seed} and \ref{fig:tas} generates only one maximal assembly, the one depicted in \RefFig{fig:assembly}.
  If the seed had a glue $c$ to the north side, then the system would not have been confluent because both tiles A and C could assemble north of \Seed and the maximal assembly would not be unique.
\end{example}

For assembly paths $\{\Assembly_i\}_{i\in I}$ we say that $\AssemblyOther=\cup_{i\in I} \Assembly_i$ if $\AssemblyOther$ has a binding graph that is union of the paths that are binding graphs for $\Assembly_i$ for all $i\in I$, and $\AssemblyOther|_{\Domain{\Assembly_i}}=\Assembly_i$. 
Note this is well defined because all paths are pairwise compatible.

\begin{lemma}
  A stable assembly in a confluent TAS \TAS is a union of assembly paths. 
\end{lemma}

\begin{proof}
  It follows directly from the fact that the binding graph of a stable assembly is connected and \TAS is confluent.
\end{proof}

\begin{corollary}
  The maximal assembly of \TAS, is $\AlphaMax=\cup_{\Assembly \in \Pi} \Assembly$ where $\Pi$ denotes the set of all assembly paths.
\end{corollary}

In the rest of the paper, to simplify notations, \AlphaMax is often used to refer to its binding graph.
In particular, $\Path\SubGraphOf\AlphaMax$ denotes that \Path is a subgraph of the binding graph of $\AlphaMax$.

\paragraph{Non-causal.}
A vertex $B$ in $\IntegerSet^2$ is \emph{non-causal} for vertex $A$ in $\IntegerSet^2$ if either $A$ does not belong to \AlphaMax or there is a path in \AlphaMax from \ZZOrigin to $A$ that does not contain $B$.
The set of non-causal vertices for $A$ in
$\Domain{\AlphaMax}$ is
\begin{equation*}
  \NonCausal{A}=\left\{
    B\in 
    \IntegerSet^2
    \mid
    \exists m\in\DirectionSet^{*},
    \ZZOrigin.m\SubGraphOf\AlphaMax
    \text{ with } \ZZOrigin+\Vect{m}=A 
    \text{ and } B\not\in \Domain{\ZZOrigin.m}
  \right\}
  \cup\{A\}
\end{equation*}
The point $A$ itself is included in the set in order to simplify later expressions. 
By the definition, if $A\not \in \Domain{\AlphaMax}$ then \NonCausal{A} is whole $\IntegerSet^2$.

An assembly $\Assembly$ that has a path as a binding graph may not necessarily be part of an assembly path, i.e., may not necessarily be part of a producible assembly.
If $\Assembly$ starts with a tile $t_0$, it may be the case that every assembly path (that starts from the seed, i.e., the origin) that reaches $t_0$ also forms an obstacle for generating $\Assembly$ starting at $t_0$.
The non-causal set of vertices for a given point in $x\in \IntegerSet^2$ identifies the region that is free from those obstacles; in other words, $B$ is in the non-causal region for $A\in\Domain{\AlphaMax}$ if one can reach $A$ within $\AlphaMax$ without passing through $B$. In particular, we have the lemma. 

\begin{lemma}
  \label{non-causal-extension}%
  Let $A\in\Domain{\AlphaMax}$.
  Suppose $p$ is a free path, and $C.p$ is a path such that there is an assembly $\Assembly$ (not necessarily producible) whose binding graph has $C.p$ as a subgraph and $\Domain{\Assembly}=\Domain{C.p}$ with a start tile $\Assembly(C)=\AlphaMax(A)$.
  If $\Domain{A.p}\subseteq \NonCausal{A}$ then $A.p\SubGraphOf\AlphaMax$.
\end{lemma}

\begin{proof} 
  Inductively, for $p=\epsilon$ we have $A.p=A\SubGraphOf\AlphaMax$. Suppose there is an assembly $\Assembly$ with binding graph
  spanned by the path $C.p'=C.pd$ such that 
  $\Domain{A.pd}\subseteq\NonCausal{A}$ and $A.p\SubGraphOf\AlphaMax$. 
  Let $B=(A+\Vect{p}).d$ be the end vertex of $A.pd$, and hence it is in $\NonCausal{A}$.
  Let $\ZZOrigin.q$ be an assembly path in $\AlphaMax$ such that $B\not \in \Domain{\ZZOrigin.q}$ and $\ZZOrigin +\Vec{q}=A$.
  Note that $q$ may use vertices in $\Domain{A.p}$ although all of the vertices along $A.p$ are in $\NonCausal{A}$.
  Suppose $p_1p_2=p$ is such that $p_1$ is the longest prefix of $p$ where $A+\Vec{p_1}\in \Domain{\ZZOrigin.q}$ (the suffix $p_2$ could be empty). If $p_1=\epsilon$ then trivially $\ZZOrigin.qpd\SubGraphOf\AlphaMax$. 
  Let $q'$ be the prefix of $q$ such that $A+\Vec{p_1}=\ZZOrigin+\Vec{q'}$.
  Then, $\ZZOrigin.q'p_2d$ is a subgraph of a binding graph of a producible assembly because \TAS is confluent, $\ZZOrigin.q'p_2$ is a path in \AlphaMax, and $p_2d$ is a suffix of an assembly path that exists.
  Therefore, the edge between $A+\Vect{p}$ and $B$ exists, i.e., $A.pd\SubGraphOf\AlphaMax$.
\end{proof}

The above lemma shows that if $A,C\in \Domain{\AlphaMax}$ and $\AlphaMax(A)=\AlphaMax(C)$, for every path $p$, if $C.p\SubGraphOf\AlphaMax$ then $A.p\SubGraphOf\AlphaMax$ as soon as $\Domain{A.p}\subseteq\NonCausal{A}$. In particular, for an assembly path \Path, for any positive index $i$, \PathFrom{\Path}{i} and \PathAfter{\Path}{i} are both in \NonCausal{\Path_{i}}.
Moreover, the site in $\IntegerSet^2$ that is not in \NonCausal{\Path_{i}} must be in \PathBefore{\Path}{i} (and \PathTo{\Path}{i}).

The lemma below shows that $\hat f=\CoGrow{b}{f}{b'}{f'}$ allows to extend paths within $\AlphaMax$ in a compatible way since \TAS is confluent. We use the following setup.

Let $bqf$ and $b'q'f'$ be two bi-infinite free paths with $b$, $b'$ being backward infinite, $q$ and $q'$ finite and starting with the same direction and $f$ and $f'$ forward infinite.
Let $\hat q\hat f=\CoGrow{b}{qf}{b'}{q'f'}$ be defined and $\hat q$ is the maximal portion of the co-growth that consists of segments of $q$ or $q'$ only.
Let $A,A'\in \IntegerSet^2$ and consider the right regions $R,R'$ of $b.A.qf$, and $b'.A'.q'f'$, respectively. We further suppose that $A.q\SubGraphOf \AlphaMax$ and $A'.q'\SubGraphOf \AlphaMax$ and $\AlphaMax(A)=\AlphaMax(A')$. 

\begin{lemma}[co-Grow compatibility]
  \label{lem:co-grow-tile}
  If $R\subseteq\NonCausal{A}$ and $R'\subseteq\NonCausal{A'}$ then both paths $A.\hat q$ and $A'.\hat q$ are subgraphs of \AlphaMax.
\end{lemma}

\begin{proof} Because $A.\hat q$ is in $R\cap R'$,
  the hypotheses imply that $\Domain{A.\hat q}$ is in \NonCausal{A} and $\Domain{A'.\hat q}$ in \NonCausal{A'}, hence by \RefLem{non-causal-extension} the paths $A.\hat q$ and $A'.\hat q$ are subgraphs of $\AlphaMax$.
  Furthermore, whenever the paths $q$ and $q'$ intersect, because of the confluence of the \TAS, the tiles in \AlphaMax corresponding to these intersections must coincide. Hence $A.\hat q$ and $A'.\hat q$ are binding graphs of the same assembly in $\TAS$; just one is a translation of the other. 
\end{proof}


\section{\AlphaMax has an ultimately periodic assembly path}
\label{sec:nice}

In this section we show that an infinite $\AlphaMax$ in a confluent tiling assembly system must contain an ultimately periodic path.
There are two cases:
\begin{enumerate}[label=(\arabic*)]
\item\label{nice:case1}
  \AlphaMax has an infinite assembly path having an empty intersection with an ultimately periodic path in the grid $\IntegerSet^2$, and 
\item\label{nice:case2}
  all infinite assembly paths in \AlphaMax intersect all ultimately periodic paths in $\IntegerSet^2$ infinitely often. 
\end{enumerate}
We prove that the first case is equivalent with \AlphaMax having an ultimately periodic assembly path, and the second case is impossible. 
We start with \RefLem{lem:nice} saying that the existence of an ultimately periodic path in \AlphaMax is equivalent to case\,\ref{nice:case1}.

Then, for a given $\delta>0$, we consider a finite segment $\Path_{\IntegerInterval{\ell}{\,r}}$ (called `off-the-wall') of an assembly path \Path such that all intersections of 
$\Path_{\IntegerInterval{\ell}{\,r}}$ with the $x$-axis ($y=0$) lie within a finite segment of the $x$-axis
$y=0$ of length $\delta$.
We consider the area `above' $y=0$, bounded by the $x$-axis and off-the-wall path $\Path_{\IntegerInterval{\ell}{\,r}}$. 
We observe (\RefLem{lem:OtW-unbounded-height}) that if, for a given $\delta$, the set of such areas above $y=0$ in \AlphaMax is not bounded, then the property\,\ref{lem:nice:avoid} of \RefLem{lem:nice} holds and \AlphaMax satisfies case\,\ref{nice:case1}.

Finally, in \RefTh{th:nice}, considering case\,\ref{nice:case2}, we start with the assumption that any infinite assembly path in \AlphaMax intersects the $x$-axis infinitely often.
This provides infinitely many off-the-wall assembly paths such that their height cannot be bounded. We show that there is $\delta$ and 
a set of off-the-wall paths above $\delta$, that are sufficiently high such that it is possible to use co-grow to prove that the set of areas for this $\delta$ cannot be bounded.
This proves the condition of \RefLem{lem:OtW-unbounded-height}, and consequently \RefLem{lem:nice} holds, contradicting possibility of case\,\ref{nice:case2}.

\begin{lemma}%
  \label{lem:nice}%
  The following two properties are equivalent:
  \begin{enumerate}
  \item\label{lem:nice:path}
    \AlphaMax contains an ultimately periodic assembly path; and
  \item\label{lem:nice:avoid}
    there is a point $A$ in $\IntegerSet^2$, an infinite periodic free path $p^{\omega}$ and an infinite path $\ZZOrigin.\Path\SubGraphOf \AlphaMax$ with $\Domain{A.p^\omega}\cap \Domain{\ZZOrigin.\Path} =\emptyset$.
  \end{enumerate}
\end{lemma}

\begin{proof}
  \RefPropertyEnum{lem:nice:path} implies \RefPropertyEnum{lem:nice:avoid} by taking the ultimately periodic path as \Path, the same period $p$ and choosing a point $A$ sufficiently away from the seed.
  It remains to prove that \RefPropertyEnum{lem:nice:avoid} implies \RefPropertyEnum{lem:nice:path}.

  First, we consider the case that $p$ extends eastwards (i.e., \Vect{p} has a positive first component) and that the infinite path \Path infinitely extends eastwards north of $A.p^{\omega}$ (i.e., for all large enough $x_0$, the vertical line $x=x_0$ intersects both \Path and $A.p^{\omega}$, and the intersection with \Path has a larger $y$ value than the intersection with $A.p^{\omega}$).
  By replacing, if necessary, $A$ by $A+k\Vect{p}$ for some positive $k$, we also assume that the vertical line passing through $A$ intersects \Path.
  Let $a$ be the least positive integer such that $A+a\Vect{\North}$ is $\Path_i$ for some $i$.
  The situation is depicted in \RefFig{fig:nice-eq:a}.

  \begin{figure}[hbt]
    \centerline{%
      \subcaptionbox{\label{fig:nice-eq:a}%
        general setting}{%
        \includegraphics{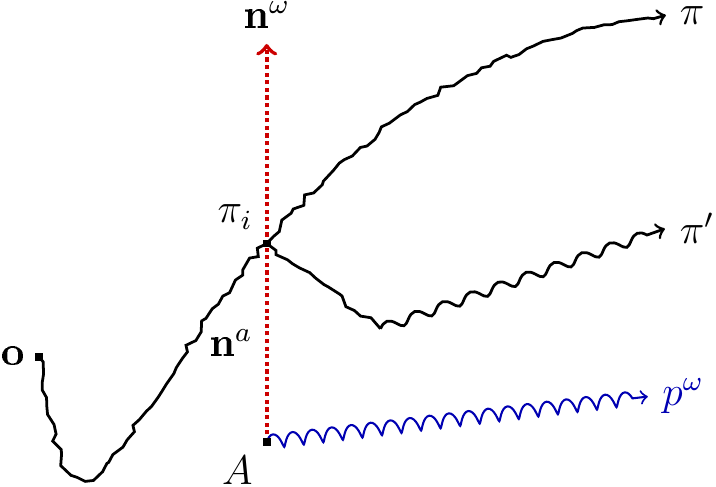}%
      }%
      \qquad%
      \subcaptionbox{\label{fig:causal}%
        causality issue}{%
        \includegraphics{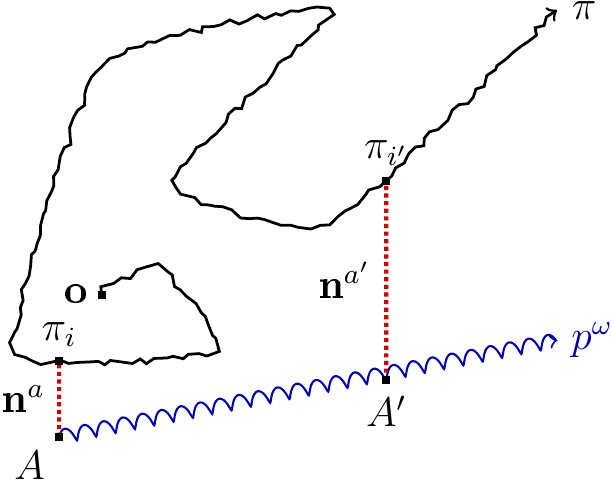}%
      }
    }
    \caption{Situation for \RefLem{lem:nice}.}
    \label{fig:nice-eq}
  \end{figure}

  The path $\PathOther=^{\omega}\Reverse{p}.A.\North^{a}.\PathFrom{\Path}{i}$ is bi-infinite.
  By hypothesis on $A.p^{\omega}$ and definition of $\Path_{i}$ the finite path $\Path_{\IntegerInterval{0}{i-1}}$ does not cross the infinite path \PathOther.
  So, $\Path_{\IntegerInterval{0}{i-1}}$ is in the interior of either the right, or the left region of \PathOther. 
  This means that the right region of \PathOther is included in \NonCausal{\Path_{i}} if and only if \ZZOrigin belongs to the left region.
  
  If \ZZOrigin is in the right region of \PathOther, that is $\ZZOrigin$ is not included in \NonCausal{\Path_{i}}, then $\Path_{\IntegerInterval{0}{i-1}}$ is inside the region as illustrated in \RefFig{fig:causal}.
  In this case, let $A'=A+k\Vect{p}$ for some positive $k$ and let $\Path_{i'}$ be the intersection of $\Path$ and $A'.\North^\omega$.
  By choosing $k$ large enough, this intersection cannot be on the finite path $\Path_{\IntegerInterval{0}{i-1}}$ and, thus, $i<i'$.
  By construction $\Path_{i}$ is in the left region of $^{\omega}\Reverse{{p}}.A'.\North^{a'}.\PathFrom{\Path}{i'}$.
  Since $i<i'$ then the whole path $\Path_{\IntegerInterval{0}{i'}}$ has to be in this left region and so is \ZZOrigin.
  Thus, the right region of $^{\omega}\Reverse{{p}}.A'.\North^{a'}.\PathFrom{\Path}{i'}$ is included in \NonCausal{\Path_{i'}}.
  From now on,  we can assume w.l.o.g. that $A$ and $\Path_i$ are such that the right region of \PathOther is included in \NonCausal{\Path_{i}}.

  Let $\Pi$ be the set of forward infinite paths in \AlphaMax that start at $\Path_{i}$,  are in the right region of \PathOther, and do not intersect $A.p^{\omega}$.
  The set $\Pi$ is not empty because it contains \PathFrom{\Path}{i}.
  Let $\Path'$ be the path in $\Pi$ such that, except for $\Path'$, there is no other infinite path in $\Pi$ that is inside the right region $R^*$ of $^{\omega}\Reverse{{p}}.A.\North^{a}.\Path'$.
  This path does exist because $\IntegerSet^2$ is discrete and it can be inductively defined from $\Path_{i}$ as follows. 
  Starting with $\North.\Path_i$ let $d$ be the rightmost direction such that there is an infinite path in $\Pi$ with prefix $\Path_i.d$. We set $\Path_0'=\Path_i$ and $\Path'_1=\Path_i+d$.
  Given $k\ge 0$ and a path $\Path'_0\cdots \Path_k'$, let $d$ be the rightmost direction with respect to $\Path'_{k-1}\Path'_k$ such that there is an infinite path in $\Pi$ starting with $\Path_k'.d$ and does not intersect $A.p^\omega$.
  Then, we set $\Path_{k+1}'=\Path_k'+d$.
  We denote the right region of $^{\omega}\Reverse{{p}}.A.\North^{a}.\Path'$ by $R^*$.
  By construction $\Path'$ is the only path in $R^*$ that starts at $\Path_i$ and does not intersect $A.p^\omega$.
  
  We consider an infinite set of vertices $\Path'_{j_i}$ ($i=1,2,\ldots$) on $\Path'$ in the following way.
  Let $k$ be minimal such that $(A+k\North).p^\omega \cap \Path'\not = \emptyset$ (see \RefFig{fig:outer-poi}). If this intersection is infinite $\{\Path'_{j_1},\Path'_{j_2},\ldots\}$ then these vertices are the desired set. If the intersection is finite, say $\{\Path'_{j_1},\Path'_{j_2},\ldots\Path'_{j_s}\}$ we set $k_1=k$ and $A'=A+k_1\North$. 
  Let $\Path'_{j_s}=A'+n\Vect{p}+\Vect{p'}$ for some proper prefix $p'$ of $p$.
  Then, we set $A_1=A'+(n+1)\Vect{p}=\Path_{j_s}+\Vect{q_1}$ where $q_1$ is a suffix of $p$ such that $p'q_1=p$. Consider the minimal $k_2$ such that $(A_1+k_2\North).p^\omega\cap \PathFrom{\Path'}{j_s}\not = \emptyset$.
  If this intersection is infinite, we append the infinite sequence of vertices to $\{\Path'_{j_1},\Path'_{j_2},\ldots\Path'_{j_s}\}$ to obtain the desired set.
  Otherwise, let $\{\Path'_{j_{s+1}},\Path'_{j_{s+2}},\ldots\Path'_{j_{t}}\}$ be the set of intersection vertices. We repeat the process by setting $A_2=\Path'_{j_{t}}+\Vect{q_2}$ where $q_2$ is a suffix of $p$ such that $A_2.p^\omega$ is a subpath of $(A_1+k_1\North).p^\omega$, and take $k_3$ to be minimal such that $(A_2+k_3\North).p^\omega\cap\PathFrom{\Path'}{j_{t}} \not = \emptyset$, etc. 
  
  Because $\Path'$ does not intersect $A.p^\omega$, by construction, for each $\Path_{j_i}$ ($i=1,2,\ldots$) there is $q_i$ suffix of $p$ such that $\Path_{j_i}.q_ip^\omega$ is a subpath of 
  $(A+k\North).p^\omega$ for some $k$ and moreover, 
  $^{\omega}{\Reverse{{p}}} \Reverse{q_i}\North.\PathFrom{\Path'}{j_i}$ is a bi-infinite path entirely in region $R^*$.

  All this is exemplified in \RefFig{fig:outer-poi}.
  The forward infinite path $A.p^{\omega}$ is shifted by $k_1\Vect{\North}$ to intersect $\Path'$ (and there is no intersection for a lesser $k$).
  There is only one intersection, $\Path'_{j_1}$ and $q_1$ is the free path from $\Path'_{j_1}$ to $A_1$.
  The process is restarted from $A_1$ instead of $A$.
  This leads to two intersection points $\Path'_{j_2}$ and $\Path'_{j_3}$ with respective free paths $q_2$ and $q_3$.
  The process then restarts from $A_2$.

  \begin{figure}[hbt]
    \centerline{%
      \includegraphics[scale=.75]{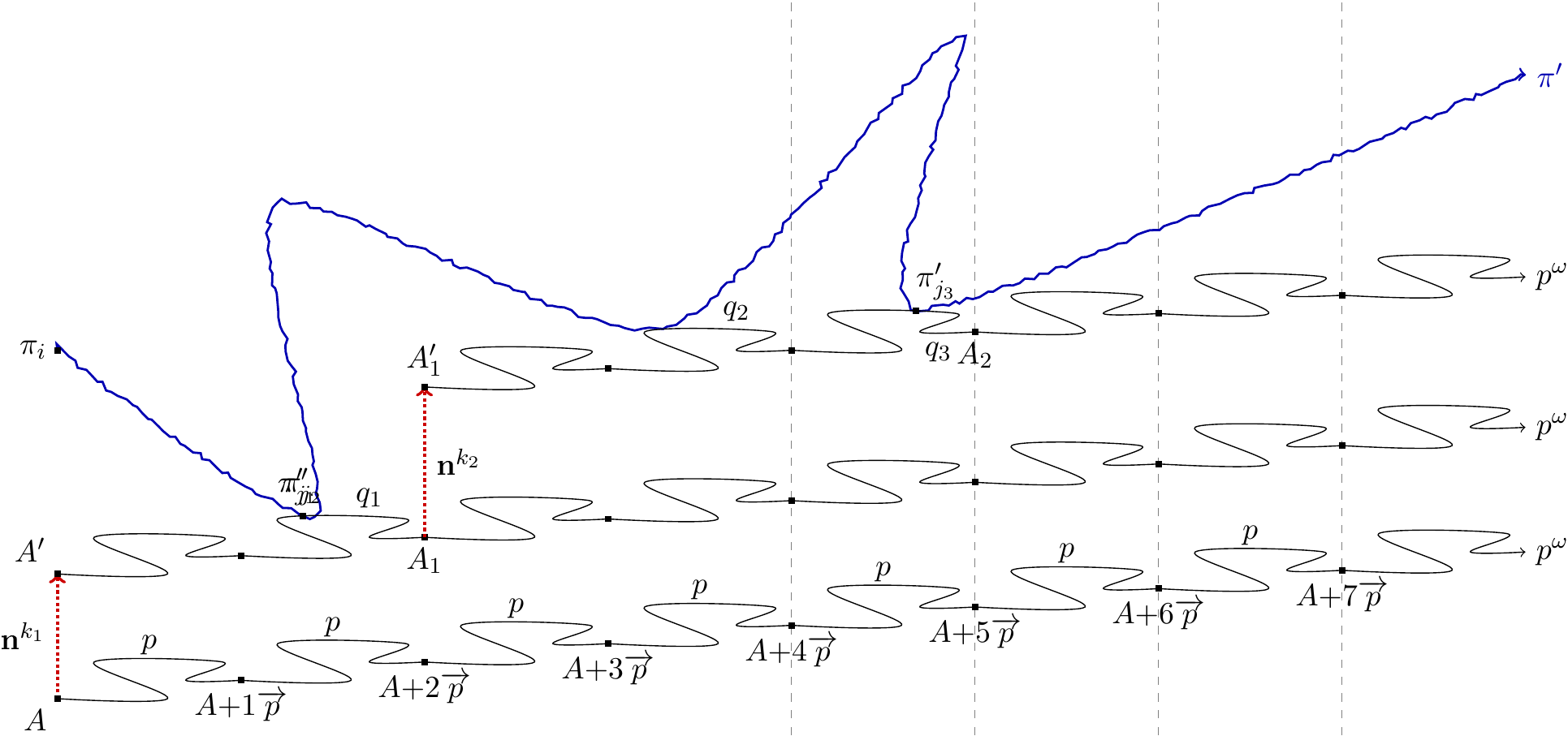}%
    }
    \caption{Extraction of special points of $\Path'$.}
    \label{fig:outer-poi}
  \end{figure}

  Since there are infinitely many such $\Path'_{j_i}$, there must be two distinct indices $j_1$, $j_2$ ($j_1<j_2$) such that $\AlphaMax(\Path'_{j_1}) =\AlphaMax(\Path'_{j_2})$, $\Path'$ follow $\Path'_{j_1}$ and $\Path'_{j_2}$ in the same direction and $q_{j_1}=q_{j_2}$ ($=q$).
  Since $^{\omega}{\Reverse{p}}\Reverse{q}\North.\PathFrom{\Path'}{j_1}$ and $^{\omega}{\Reverse{p}}\Reverse{q}\North.\PathFrom{\Path'}{j_2}$ are bi-infinite paths, the forward free path $f$ generated by
  $\CoGrow{^{\omega}{\Reverse{p}}\Reverse{q}\North}{\PathFrom{\Path'}{j_1}}{^{\omega}{\Reverse{p}}\Reverse{q}\North}{\PathFrom{\Path'}{j_2}}$ is infinite since the right regions of both
  $^{\omega}{\Reverse{{p}}}\Reverse{{q}}\North.\ZZOrigin.\PathFrom{\Path'}{j_1}$ and
  $^{\omega}{\Reverse{{p}}}\Reverse{{q}}\North.\ZZOrigin.\PathFrom{\Path'}{j_2}$ contain the backward infinite path $^{\omega}{\Reverse{{p}}}\Reverse{{q}}\North.\ZZOrigin$ (\RefLem{inf-co-grow}).
  
  Moreover, the right region of $^{\omega}{\Reverse{{p}}}\Reverse{{q}}\North.\PathFrom{\Path'}{j_1}$ is included in $R^*$ and hence is in \NonCausal{\Path_{j_1}} because the \ZZOrigin is outside this region.
  Then, by \RefLem{lem:co-grow-tile}, the path $\Path'_{j_1}.f$ belongs to \AlphaMax.

  By definition of $\Path'$, $\Path'$ is the only path in $\Pi$ that is inside region $R^*$, and because 
  $\Path'_{\IntegerInterval{i}{j_1}}.f$ is in $\Pi$, it must be included in the left region of $^{\omega}\Reverse{{p}}.A.\North^{a}.\PathFrom{\Path'}{i}$.
  By co-grow, ${\Path'}_{j_1}.f$ also has to be in the right region $R^*$ which includes the right regions of $^{\omega}{\Reverse{p}}\Reverse{q}\North.\PathFrom{\Path'}{j_1}$ and $^{\omega}{\Reverse{p}}\Reverse{q}\North.\PathFrom{\Path'}{j_2}$.
  Therefore ${\Path'}_{j_1}.f$ has to be on the boundary of $R^*$ and hence $\Path'={\Path'}_{\IntegerInterval{1}{j_1}}.f$.
  The same is true for the index $j_2$; so $\Path'={\Path'}_{\IntegerInterval{1}{j_1}}.f={\Path'}_{\IntegerInterval{1}{j_2}}.f$.
  Since ${j_1}< j_2$, $\Path'$ has to be ultimately periodic and thus, by \RefLem{path-assembly}
  there is an ultimately periodic assembly path in \AlphaMax with domain $\Path'$.

  \smallskip
  
  If $p$ and $\Path$ do not extend eastwards, the TAS \TAS can be rotated until they do so.
  A TAS \TAS is rotated (by 90° clockwards) by rotating its seed and all its tile types.
  A free path is rotated by replacing \North by \East, \East by \South…
  TAS \TAS can be north-south symmetrized in a similar way.
  The obtained TAS is also confluent and generate the same \AlphaMax up to rotations and up-down symmetry.
  So without loss of generality, we consider that $p$ extends eastwards.
  If \Path extends infinitely eastwards, but not north of $A.p^{\omega}$, then it does so after a north-south symmetry of $p$ and \TAS.

  If \Path does not extends infinitely eastwards, then there is some $A'$ in $\IntegerSet^2$ such that \Path does not intersect $A'.\North^{\omega}$.
  Then, we consider $A'$ and $p=\North$ and a rotation of \TAS.
  If \Path extended infinitely northwards, then, after rotation (of $p$ and \TAS), \Path extends eastwards and $p$ is \East.
  Otherwise we can perform a rotation again.
  Because the path \Path is infinite, it has to extend in at least one of the four basic directions; so eventually the infinite path \Path infinitely extends eastwards north of $A.p^{\omega}$.
  
  This concludes the proof.
\end{proof}

\begin{figure}[hbt]
  \centering
  \qquad%
  \qquad%
  \includegraphics{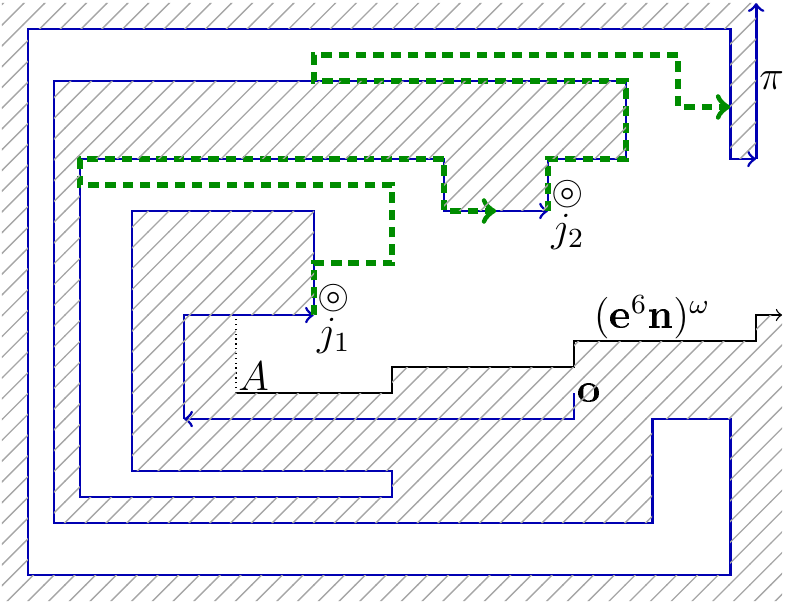}%
  \qquad%
  \qquad%
  \qquad%

  \caption{\label{fig:nice-eq:b}%
    path \Path (blue) with dashed (green) co-grow. Here $p=\East^6\North$.
    The shaded area is the left region that may contain other paths.
    The white area is the right region where paths can co-grow.}
\end{figure}

Observe that the path $\Path$ can be quite complicated, as depicted in \RefFig{fig:nice-eq:b}.
In the figure $p=\East^6\North$ and the unshaded (white) region is the right region of $^{\omega}\Reverse{{p}}.A.\North^{a}.\Path$.
This region belongs to $\NonCausal{A+a\North}$. 
An example of co-grow is displayed with dashed lines showing that $\Path_{\IntegerInterval{i}{+\infty}}$ cannot be the path $\Path'$ defined above.

Note that the above lemma does not provide a construction of the ultimately periodic path $\Path'$ but only shows its existence. In 
\citep{durand-lose+hoogeboom+jonoska19arxiv}
we show an algorithmic way how to use \RefLem{lem:nice} to provide a finite description of \AlphaMax that provides ultimately periodic paths comprising \AlphaMax. 

\begin{corollary}%
  \label{cor:new-periodic-path}
  Let $\Path$ be an infinite path in \AlphaMax, $mp^\omega$ be a free path and $i$ be a positive number such that $\Domain{\ZZOrigin.\Path}\cap \Domain{\Path_i.mp^\omega}=\{\Path_i\}$.
  Then, one of the three following possibilities appears:
  \begin{enumerate}
  \item $\Path$ is ultimately periodic, 
  \item there is an ultimately periodic path in \AlphaMax intersecting $\Domain{\Path}$ on an infinite set, or
  \item there is an ultimately periodic path in \AlphaMax strictly inside the right region of $\Reverse{{ mp^\omega}}.\Path_{\IntegerInterval{i}{\infty}}$.
  \end{enumerate}
\end{corollary}

\begin{proof}
  The proof follows directly from the proof of \RefLem{lem:nice}. If the first two conditions are not satisfied, then the path $\Path'$ constructed in the proof of \RefLem{lem:nice} satisfies the third condition.
\end{proof}

\begin{definition}[off-the-wall path]
  Let $\delta$ a positive integer. 
  A finite path $\Path=\Path_0\cdots\Path_r$ is \emph{(rightwards) $\delta$-off-the-wall} if there exists a positive integer $\ell$ less than $r$ such that: 
  \begin{enumerate}
  \item \Path is an assembly path in \AlphaMax, 
  \item there is $x_0\in \IntegerSet$ satisfying 
    $\Path_{\ell}=(x_{0},0)$
    and
    $\Path_{r}=(x_{0}+\delta,0)$,
  \item 
    $ \Domain{\Path}\ \cap\ \Domain{{^{\omega}\East}.\Path_{r}} = \Domain{\Path}\ \cap\ \Domain{\Path_{\ell}.\East^{\omega}}$,
    and
  \item and \ZZOrigin is in the right region of the bi-infinite path
    \begin{equation}
      \label{eq:nice:bi-infinite}
      ^{\omega}\East
      .\Path_{\IntegerInterval{\ell}{\,r}}
      .\East^{\omega}
      \enspace.
    \end{equation}
  \end{enumerate}
\end{definition}

The \emph{leftwards $\delta$-off-the-wall} is defined in a symmetric way by swapping east and west.
We use simply the phrase $\delta$-off-the-wall to denote rightwards $\delta$-off-the-wall, unless otherwise stated. 

\begin{figure}[hbt]
  \centering%
  \includegraphics{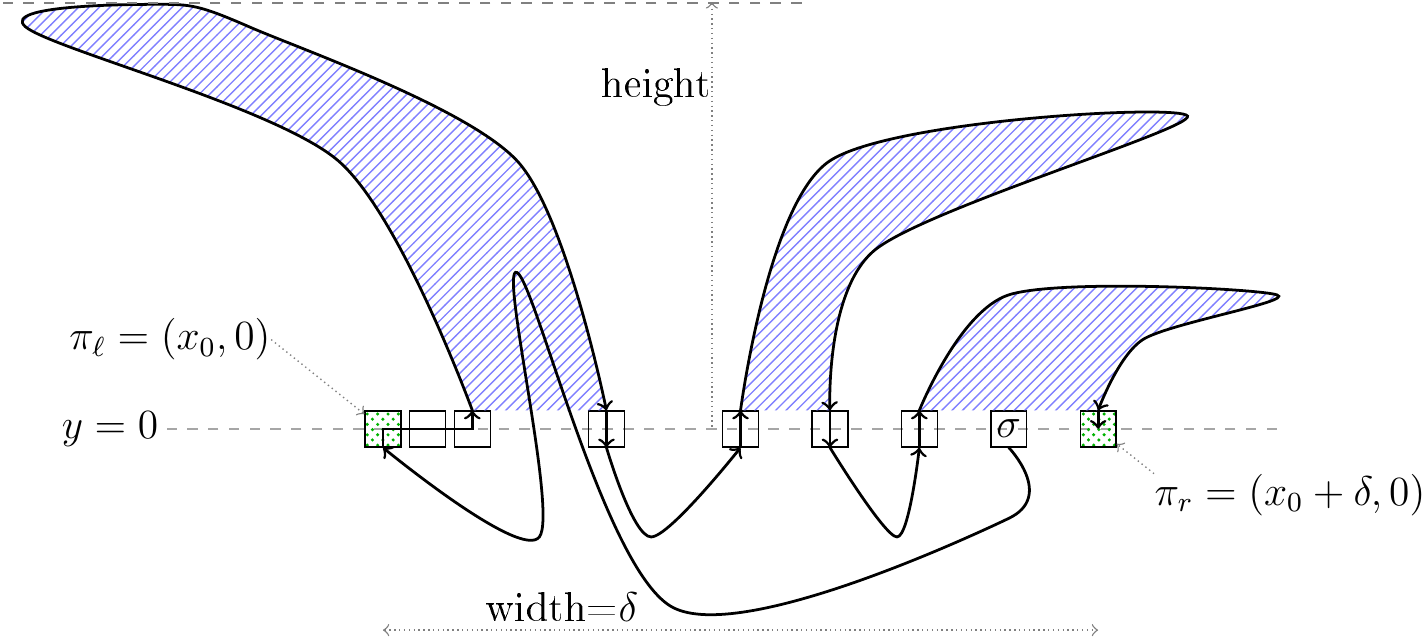}%
  \caption{$\delta$-off-the-wall path (the surface above is shaded).}
  \label{fig:off-the-wall}
\end{figure}

The positive number $\delta$ is called the \emph{width} of the off-the-wall path.
The \emph{wall} is the subgraph of $\IntegerSet^2$: $ ^{\omega}\East.\ZZOrigin.\East^{\omega}$.
The segment $\Path_{\IntegerInterval{\ell}{\,r}}$ of \Path is called \emph{above-the-wall} part of \Path (even though some of its portions might be 
under `the wall' as illustrated in \RefFig{fig:off-the-wall}).
We call $\Path_\ell$ the left end, and $\Path_r$ the right end of the off-the-wall path.

\begin{definition}[height, surface and area above the wall]
  Let \Path be any $\delta$-off-the-wall path.
  Its {\em height} is the maximal $y$-coordinate it reaches above the wall.
  The \emph{surface above (the wall)} is the intersection of the left region of 
  $^{\omega}\East
  .\ZZOrigin.\East^
  {\omega}$
  with the right region of $^{\omega}\East
  .\Path_{\IntegerInterval{\ell}{\,r}}
  .\East^{\omega}$.
  The \emph{area above} is the area of the surface above the wall.
  If $\Path$ is an assembly path, the \emph{wall valuation} of an off-the-wall path is the pair of tile types at $\Path_{\ell}$ and $\Path_{r}$.
\end{definition}

The notions of off-the-wall path, the height above, and surface above are illustrated in \RefFig{fig:off-the-wall} where the surface above is shaded.
The area above is the area of the shaded portion.
The wall valuation corresponds to the pair of green dotted tiles.

The following lemma allows to combine and extend off-the-wall paths.

\begin{lemma}[Off-the-wall combination]
  \label{lem:off-the-wall-combination}
  Let
  $\Path=\Path_0\cdots\Path_{\ell}\cdots\Path_{r}$ and
  $\Path'=\Path_0'\cdots\Path'_{\ell'}\cdots\Path'_{r'}$ be two $\delta$-off-the-wall assembly paths with the same wall valuation.
  There exists a free path $g$ such that: the path $\Path_{\IntegerInterval{0}{\ell}}.g$ (resp. $\Path'_{\IntegerInterval{0}{\ell'}}.g$) is a $\delta$-off-the-wall path with the same wall valuation and the area of its surface above the wall contains the area of the surface above the wall of $\Path$ (resp. $\Path'$).
\end{lemma}

\begin{proof}
  The portions above the wall of \Path and $\Path'$ are $\Path_{\IntegerInterval{\ell}{\,r}}$ and $\Path'_{\IntegerInterval{\ell'}{\,r'}}$ respectively. Since they have the same 
  wall valuations, the tile type at their extreme intersections with the $x$-axis are identical, but possibly shifted horizontally. 
  By the definition of the off-the-wall path, everything in the left region of, including, $^{\omega}\East.\Path_{\IntegerInterval{\ell}{\,r}}.\East^{\omega}$ (resp. in the left region of $^{\omega}\East.\Path'_{\IntegerInterval{\ell'}{\, r'}}.\East^{\omega}$) belongs to $\NonCausal{\Path_{\ell}}$ (resp., $\NonCausal{\Path_{\ell'}'}$).
  Let $p$ and $p'$ be the free paths corresponding to $\Path_{\IntegerInterval{\ell}{\,r}}$ and $\Path'_{\IntegerInterval{\ell'}{\,r'}}$ respectively.
  We left-co-grow the bi-infinite paths $^{\omega}\East p\,\East^{\omega}$ and $^{\omega}\East p'\East^{\omega}$.
  
  Let $f=\LCoGrow{^{\omega}\East}{p\,\East^{\omega}}{^{\omega}\East}{p'\East^{\omega}}$.
  The infinite path $^{\omega}\East.\ZZOrigin$ is in both left regions of $^{\omega}\East.\ZZOrigin.p\,\East^{\omega}$ and $^{\omega}\East.\ZZOrigin.p'\East^{\omega}$; so $f$ is infinite by the left-co-grow version of \RefLem{inf-co-grow}.
  By construction of co-grow, $\ZZOrigin.f$ is equal to $\ZZOrigin.g\East^{\omega}$ where $g$ is a prefix of $f$ with displacement $\Vec{g}=(\delta,0)$ where $\delta$ is the common width of the off-the-wall paths $\Path$ and $\Path'$.
  By \RefLem{lem:co-grow-tile} both $\Path_{\IntegerInterval{0}{\ell}}.g$ and $\Path'_{\IntegerInterval{0}{\ell'}}.g$ are paths in \AlphaMax. 
  Because the co-grow free path $g$ takes the leftmost segment of the two paths, the area of the surface above $\Path$ (and resp. $\Path'$) is included in the area of the surface above $\Path_{\IntegerInterval{0}{\ell}}.g$ (resp. $\Path'_{\IntegerInterval{0}{\ell'}}.g$).
  The constructed off-the wall paths have the same width $\delta$ and wall valuation.
\end{proof}

Above lemma allows to generate paths in \AlphaMax that are compatible with respect to their wall valuations while increasing the areas of the surfaces above of the original paths.

\begin{corollary}%
  \label{cor:max-area}
  If the set of areas above $\delta$-off-the-wall paths with the same wall valuation is bounded then there is a unique $\delta$-off-the-wall path $\hat\Path$ with the same wall valuation whose area above the wall is maximal.
  Moreover, the height of $\hat\Path$ is maximal among all paths with the same wall valuation. 
\end{corollary}

\begin{proof}
  Consider the set of all $\delta$-off-the-wall paths with identical wall valuation.
  Suppose that this set is finite. If this set has only one path, then that path is $\hat\Path$.
  Consider two off-the-wall paths $\Path_{\IntegerInterval{\ell}{\,r}}$ and $\Path'_{\IntegerInterval{\ell'}{\,r'}}$ in this set with a maximal area above.
  By \RefLem{lem:off-the-wall-combination}, there is a free path $g$ such that both $\Path_{\IntegerInterval{0}{\ell}}.g$ and $\Path'_{\IntegerInterval{0}{\ell'}}.g$ are off-the-wall paths with the same wall valuation whose area above the wall contains the areas above both $\Path$ and $\Path'$.
  Since $\Path_{\IntegerInterval{\ell}{\,r}}$ and $\Path'_{\IntegerInterval{\ell'}{\,r'}}$ have maximal areas above the wall, it must be that $\Path_{\ell}.g=\Path_{\IntegerInterval{\ell}{\,r}}$ and $\Path_{\ell'}.g=\Path'_{\IntegerInterval{\ell'}{\,r'}}$. 
  We observe that $\Path_{\ell}.g$ (resp. $\Path'_{\ell'}.g$) must also have a maximal height because its area above the wall contains all areas above the wall for the set of all $\delta$-off-the-wall paths with the same wall valuation.
\end{proof}

\begin{lemma}%
  \label{lem:OtW-unbounded-height}
  If the set of areas above $\delta$-off-the-wall paths in $\AlphaMax$ with a given wall valuation is not bounded, then $\AlphaMax$ contains an ultimately periodic assembly path.
\end{lemma}

\begin{proof}
  We consider only the off-the-wall paths that correspond to the width $\delta$ and have the same wall valuation.
  By the lemma hypothesis, for each $k$ there is a $\delta$-off-the-wall path $\Path^{k}$ such that its area above the wall is greater than $k$.
  Let $\hat{\Path}^{0}=\Path^{0}$ and
  for $k\ge 0$ let $\hat{\Path}^{k+1}$ be obtained from $\hat{\Path}^{k}$ and $\Path^{k+1}$ by the combination lemma (\RefLem{lem:off-the-wall-combination}).
  
  All the $\hat{\Path}^{k}$ pass through $\Path^{0}_{\ell}$, and $\hat\Path^{0}_{r}$. 
  By \RefLem{lem:off-the-wall-combination}, they are all $\delta$-off-the-wall with the same wall valuation and belong to \AlphaMax
  (i.e., are assembly paths).

  Consider the the subgraph $G$ of \AlphaMax that consists of the union of all off-the-wall segments of all $\hat{\Path}^{k}$. 
  This subgraph $G$ must be a subgraph of \AlphaMax (because all co-grow $\delta$-off-the-wall paths are subgraphs of \AlphaMax) and it is infinite because the areas of the surfaces above the co-grows are not bounded. 
  The graph $G$ is also connected because it is a union of co-grows of $\hat{\Path}^{k}$.
  Moreover, $G$ has no intersections with $(\hat\Path_r+(1,0)).\East^\omega$ because all off-the-wall paths that comprise $G$ intersect the $x$-axis between $\hat\Path_\ell$ and $\hat\Path_r$.
  
  By K\"onig's lemma there is an infinite path $\Path'$ in $G$ starting at $\Path^{0}_{\ell}$.
  Because of the co-grow constructions, this infinite $\Path'$ does not intersect $\Path^0_{\IntegerInterval{0}{\ell}}$.
  Therefore $\Path=\Path^0_{\IntegerInterval{0}{\ell}}.\Path'$ is an infinite assembly path in \AlphaMax.
  Then, $({\Path}_r^0+(1,0)).\East^{\omega}$ is a periodic path in $\IntegerSet^2$ that has no intersection with the infinite assembly path $\Path$.
  By \RefLem{lem:nice}, \AlphaMax contains an ultimately periodic assembly path.
\end{proof}

In the following we observe that the property\,\ref{lem:nice:avoid} in \RefLem{lem:nice} is always satisfied in a confluent system \TAS as soon as there exists an infinite path.
To conclude this, we show that there are always off-the-wall paths with the same wall valuation and unbounded set of areas above; therefore satisfying \RefLem{lem:OtW-unbounded-height}. 
We concentrate on paths in \AlphaMax intersecting the $x$-axis an infinite number of times.
If there exists an infinite path in \AlphaMax that does not intersect $\ZZOrigin.\East^{\omega}$ (or $\ZZOrigin.\West^{\omega}$) an infinite number of times, then by \RefLem{lem:nice}, there is an ultimately periodic assembly path in \AlphaMax.
We start by observing that the heights of the off-the-wall paths that intersect both sides of the $x$-axis (that is, $\ZZOrigin.\East^{\omega}$ and $\ZZOrigin.\West^{\omega}$), infinite number of times is unbounded. 

\begin{lemma}\label{lem:infinite-OtW}
  If there is an infinite assembly path \Path intersecting both $\ZZOrigin.\East^{\omega}$ and $\ZZOrigin.\West^{\omega}$ an infinite number of times then, up to a vertical symmetry, there exist infinitely many off-the-wall paths and the set of heights of these paths is unbounded. 
\end{lemma}

\begin{proof}
  Let $W$ be the sets of indices $\ell$ of \Path such that $\Path_\ell\in\ZZOrigin.\West^{\omega}$ and all the indices (on \Path) of vertices in the intersection of \Path and $\Path_\ell.\West^{\omega}$ are greater than $\ell$.
  Let $E$ be defined similarly on the east direction.
  Both $E$ and $W$ are infinite because \Path intersects the $x$-axis an infinite number of times on both sides.
  There are infinitely many pairs $(w,e)$ such that: $w\in W$, $e\in E$, $w<e$, $\IntegerInterval{w+1}{e-1}\cap ( E\cup W ) = \emptyset$. An example is depicted in \RefFig{fig:E+W} where the indices $240$ and $300$ form such a pair, while $(130,300)$ is not because $240\in [129,299]\cap(E\cup W)$. 

  \begin{figure}[hbt]
    \centering%
    \includegraphics[scale=.9]{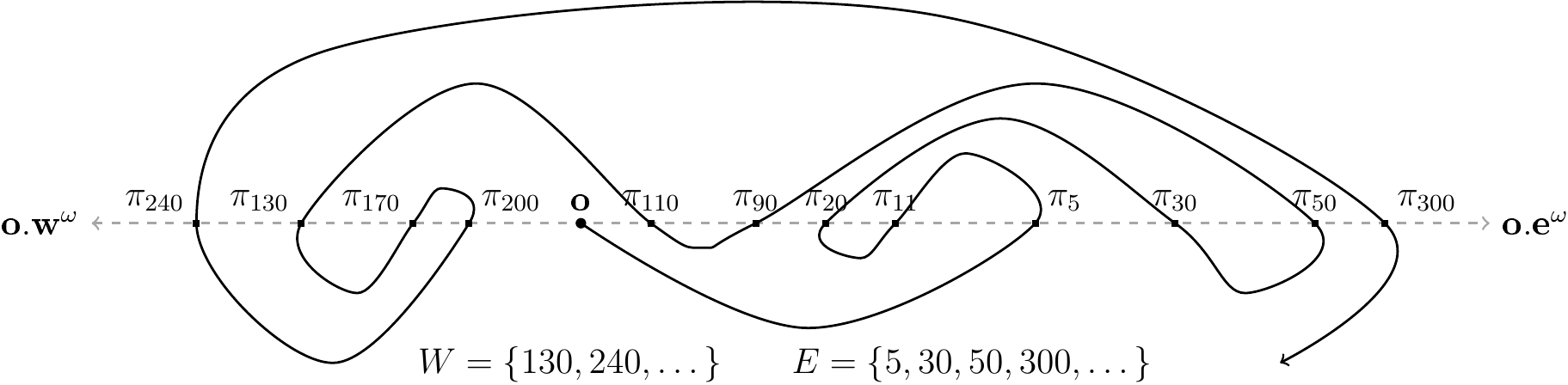}%
    \caption{Points in $E$ and $W$.}
    \label{fig:E+W}
  \end{figure}
  
  One of the left, or the right region of the bi-infinite path $^{\omega}\East.\Path_{\IntegerInterval{w}{e}}.\East^{\omega}$ must contain the entire path $\Path_{\IntegerInterval{0}{w-1}}$.
  If it is in the right region, then $\Path_{\IntegerInterval{0}{e}}$ is an off-the-wall path, otherwise, $\Path_{\IntegerInterval{0}{e}}$ is an off-the-wall path for the system that is vertically symmetric to \TAS.
  In this way we obtain infinitely many off-the-wall paths for \TAS (or for its vertically symmetric one).

  Consider two such pairs $(w,e)$ and $(w',e')$. It must be $w<e<w'<e'$ because the intervals $[w+1,e-1]$ and $[w',e']$ cannot intersect. By definition of $E$ and $W$, $\Path_{e'}$ is further to the east than $\Path_e$ and $\Path_{w'}$ is further to the west than $\Path_w$.
  Then, $\Path_{\IntegerInterval{0}{e'}}$ must have a greater height (goes `above') than the height of $\Path_{\IntegerInterval{0}{e}}$.
  As pairs of indices $(w,e)$ increase, the off-the-wall paths have to pass one `above' the other with an increasing height.
  Therefore the set of heights and the set of areas above for these paths are not bounded.
\end{proof}

The points of interest defined below are used in the proof of the main theorem of this section.

A \emph{point of interest} of an off-the-wall path \Path is any point that is west-most on any horizontal line above the wall, that is, a point $\Path_{k}$ is a point of interest if \Path does not intersect $(\Path_{k}+\Vect{\West}).\West^{\omega}$.

In particular, \PathAfter{\Path}{k} and \PathBefore{\Path}{k} do not intersect $\Path_{k}.\West^{\omega}$.
There is a point of interest on every horizontal line above the wall up to the height of \Path, and distinct points of interest are on a distinct height above the wall.
We point out that points of interest defined above have similar flavor as the notion of `visible glues' used in \citep{meunier+woods17stoc,meunier+regnault+woods20stoc,meunier+regnault+woods20archive}, except, we are not concerned with the glues
of the tiles but the vertices where they appear.
\RefFigure{fig:3-PoI} shows points of interest $\Path_{k_1}$ and $\Path_{k_2}$.
The label "no past, no future" indicates that $\Path$ does not intersect the horizontal lines to the west. 

We will use the following lemma in the main theorem.

\begin{theorem}
  \label{th:nice}
  A confluent tiling system \TAS either has a finite \AlphaMax or \AlphaMax has an ultimately periodic assembly path.
\end{theorem}

\begin{proof}
  Suppose \AlphaMax is infinite.
  If property of \RefLem{lem:nice} hold, then \AlphaMax contains an ultimately periodic path.
  So assume that the properties of \RefLem{lem:nice} do not hold.
  Hence, every infinite path in \AlphaMax is intersecting $\ZZOrigin.\East^{\omega}$ and $\ZZOrigin.\West^{\omega}$ infinitely number of times.
  Moreover, the conditions of \RefLem{lem:OtW-unbounded-height}
  do not hold, because otherwise, as seen in the proof, the conditions of \RefLem{lem:nice} hold, and there is an ultimately periodic assembly path in \AlphaMax. 
  Thus, for every width $\delta$ and every wall valuation, the set of areas above the wall of $\delta$-off-the-wall paths with the same valuation is bounded. 
  On the other side, by \RefLem{lem:infinite-OtW}, there are off-the-wall paths with heights larger than $h$ for every $h$.

  Consider $h$ large enough such that any off-the-wall path $\Path'$ with height larger than $h$ must have two distinct points of interest such that the tile types at these vertices in \AlphaMax are the same, and the edges in \Path' incident to these vertices are in the same directions. 
  Because $h$ can be arbitrarily large and $\TAS$ is finite, such $h$, off-the-wall path $\Path'$ of height $h$, and points of interest exist. 
  Consider an off-the wall path $\Path$ that has maximal area above the wall with the same wall valuation as the path $\Path'$ of height $h$.
  Then, $\Path$ is of height at least $h$ (by \RefCor{cor:max-area}) and hence, it has points of interest $\Path_{k_1}$ and $\Path_{k_2}$ (with $k_1<k_2$) on \Path such that $\AlphaMax(\Path_{k_1})=\AlphaMax(\Path_{k_2})$ and the edges $\Path_{k_1}\Path_{k_1+1}$ and $\Path_{k_2}\Path_{k_2+1}$ are in the same direction.
  
  We show that either there is an ultimately periodic assembly path in \AlphaMax with the same prefix as \Path, or there is an off-the-wall path with same wall valuation as \Path and a larger surface area above the wall (hence contradicting the maximality of the area bounded by \Path, implying that \RefLem{lem:OtW-unbounded-height} must hold, and hence there is an ultimately periodic assembly path).
  Let $m$ and $p$ be free paths that correspond to $\Path_{\IntegerInterval{0}{k_1}}$ and $\Path_{\IntegerInterval{k_1}{k_2}}$ respectively.
  Then, $\Path =\ZZOrigin.mp.\PathFrom{\Path}{k_2}$\,.
  This is illustrated in \RefFig{fig:3-PoI}.

  \begin{figure}[hbt]
    \centering%
    \includegraphics{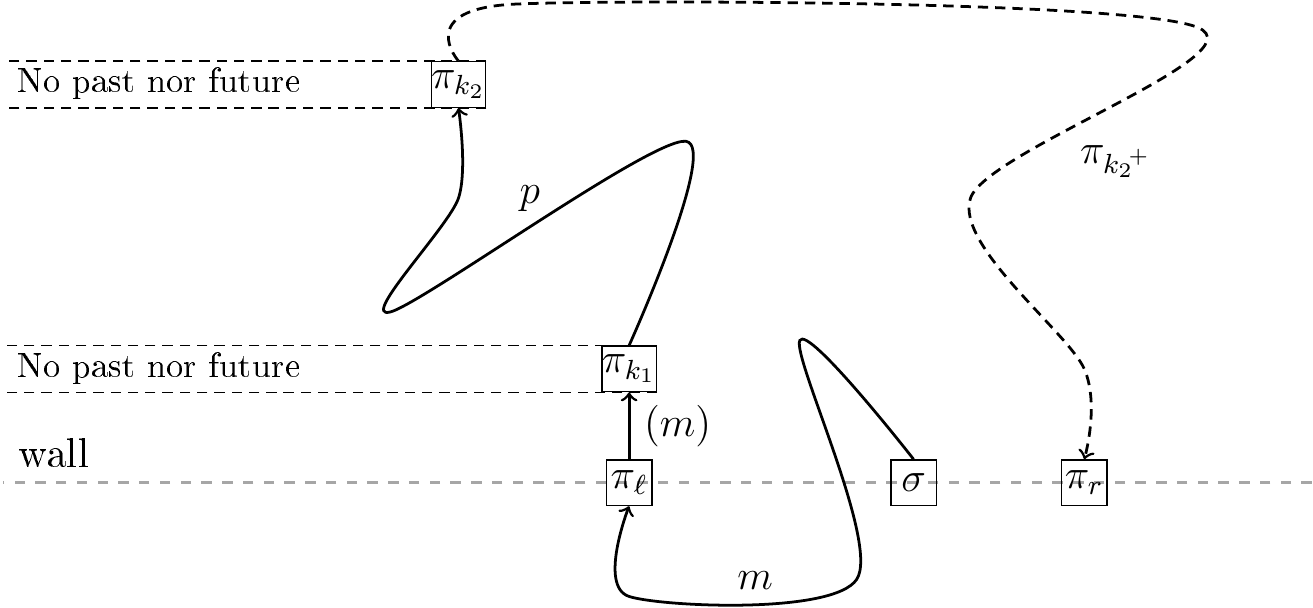}
    \caption{Two points of interest $\Path_{k_1}$ and $\Path_{k_2}$ with the same tile in \AlphaMax and the same direction following $\Path$.}
    \label{fig:3-PoI}
  \end{figure}

  Consider the paths $f_2=\Path_{\IntegerInterval{k_2}{r}}.\East^{\omega}$ and $f_1=pf_2$.
  The free paths $f_1$ and $f_2$ differ because there is at least one \North in $p$.
  Let $f$ be the \LCoGrow{^{\omega}\East}{f_1}{^{\omega}\East}{f_2}.
  Since the infinite path $^{\omega}\East.\ZZOrigin$ is in each of the left regions of $^{\omega}\East.\ZZOrigin.{f_1}$ and $^{\omega}\East.\ZZOrigin.{f_2}$, by \RefLem{inf-co-grow}, $f$ is infinite, and $f$ must end with $\East^{\omega}$. Let $f=g\East^{\omega}$ where $g$ is the segment generated by the co-grow of both segments $\Path_{\IntegerInterval{k_1}{r}}$ and $\Path_{\IntegerInterval{k_2}{r}}$.
  Because the left regions of $^{\omega}\East.\Path_{k_1}. f_1$ and $^{\omega}\East.\Path_{k_2}. f_2$ are subsets of $\NonCausal{\Path_{k_1}}$ and $\NonCausal{\Path_{k_2}}$ respectively, the free path $f$ can start from both $\Path_{k_1}$ and $\Path_{k_2}$ and both $\Path_{k_1}.g$ and $\Path_{k_2}.g$ belong to \AlphaMax by \RefLem{lem:co-grow-tile}.
  
  The infinite free path $f$ is either $f_1$, or $f_2$, or $g$ is a (leftmost) combination of $p.\PathFrom{\Path}{k_2}$ and $\PathFrom{\Path}{k_2}$, taking segments of both paths.
  In the last case when $g$ is a leftmost combination of both $p.\PathFrom{\Path}{k_2}$ and $\PathFrom{\Path}{k_2}$, taking segments of both paths, we consider the assembly paths $\Path_{k_1}.g$ and $\Path_{k_2}.g$.
  At least one of these assemblies goes strictly to the left of \Path, that is, disconnects from \Path, and then reconnects with it before the end of $g$ (otherwise $g$ is either prefix of $f_1$ or $f_2$). 
  As depicted in \RefFig{fig:enlarge}, this forms an off-the-wall path that has strictly larger area above the wall than \Path without changing the wall valuation. This is not possible because $\Path$ is supposed to have a maximal area above the wall.

  \begin{figure}[hbt]
    \centering%
    \small%
    \subcaptionbox{\label{fig:enlarge}%
      area above the wall enlargement (dotted area) when $g$ is a combination of $\Path_{\IntegerInterval{k_1}{r}}$ and $\Path_{\IntegerInterval{k_2}{r}}$}{%
      \includegraphics[scale=.7]{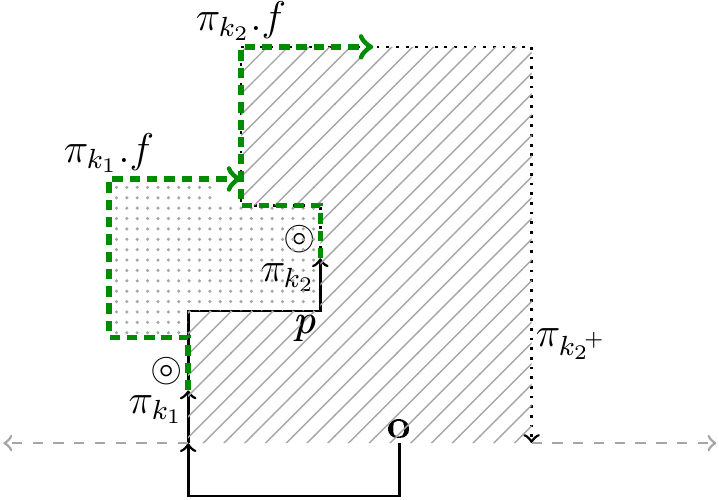}}
    \quad
    \subcaptionbox{impossible $f=f_2$%
      \label{fig:impossible:f2}}{%
      \includegraphics[scale=.7]{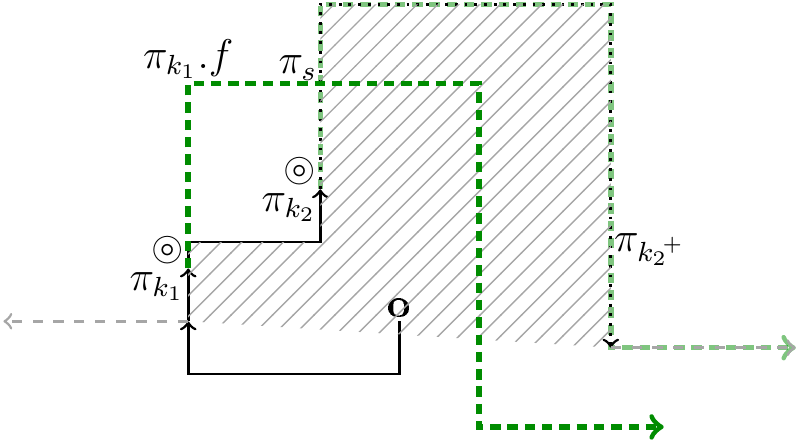}}%
    \quad
    \subcaptionbox{\label{fig:infinitely-pumpable}%
      infinitely periodic in \AlphaMax when $f=f_1$}{%
      \includegraphics[scale=.7]{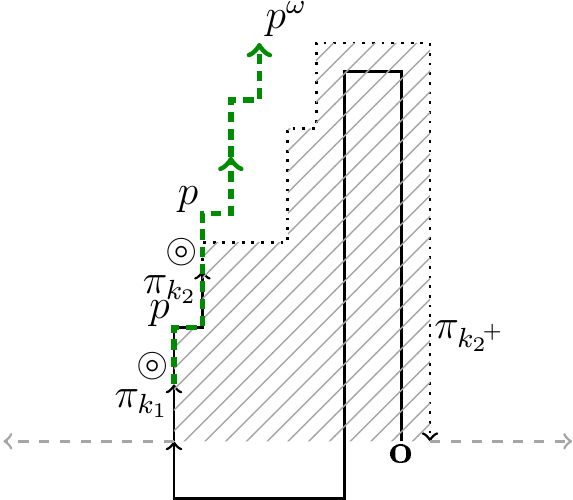}}%
    \caption{Co-grow cases for $f$.}
    \label{fig:simple-cases}
  \end{figure}

  It is not possible that $g$ is $\Path_{\IntegerInterval{k_2}{r}}$, i.e., $f$ is $f_2$, because $\Path_{k_1}.g\East^{\omega}=\Path_{k_1}.f=\Path_{k_1}.f_2$ (as illustrated in \RefFig{fig:impossible:f2}) would intersect $\Path_{k_2}.f_2$ being the same segment of \Path that starts from a point, $\Path_{k_1}$, shifted south. But then $\Path_{k_1}.f$ cannot pass north of a north shifted version of itself, and $\Path_{k_1}.f_2$ must intersect $\Path_{k_2}.f_2$. This implies that 
  $g$ is a (leftmost) combination of $p.\PathFrom{\Path}{k_2}$ and $\PathFrom{\Path}{k_2}$, taking segments of both paths, which, as seen above, produces a path with larger area above the wall than \Path and contradicts our choice of \Path.
  
  The only case left to consider is $f=f_1$, i.e., $g=\FreePath{\Path_{\IntegerInterval{k_1}{r}}}$. In this case $\Path_{k_2}.f$ is a shift of $\Path_{k_1}.f_1$ northwise and there are no vertices of $\Path_{k_2}.f_2$ strictly 
  to the left of $\Path_{k_2}.f=\Path_{k_2}.f_1$.
  Thus, $p$ is a prefix of $g$ and $\Path_{k_2}.p$ is in \AlphaMax.
  Since $\Path_{k_1}+\Vect{p}=\Path_{k_2}$, $\Path_{k_1}.p^2$ is also in \AlphaMax and, by \RefLem{lem:double-pumpable}, $p$ can form a forward infinite periodic path $p^\omega$.
  Let $A=\Path_{k_1}$ and $B=\Path_{k_2}$;
  then $A.pf_2=A.f_1=A.f$, and also $A.p$ ends at $B$. We have that $B.f_2$ is in the right region of $^\omega\East p.B.pf_2= {}^\omega\East.A.ppf_2$ because $pf_2=f$ is the left co-grow of $f_1$ and $f_2$. Similarly, $B.pf_2$ is in the right region of $^\omega\East p.B.ppf_2=\,^\omega\East.A.pppf_2$ and hence
  $B.f_2$ is in the right region of $^\omega\East.A.pppf_2$. Inductively, we have that $B.f_2$ is in the right region of $^\omega \East.A.p^\omega$, i.e., $A.p^\omega$
  is in the left region of $^\omega \East.A.f$.
  Since $\Path_{\IntegerInterval{0}{k_1-1}}$ is in the interior of the right region of $^\omega \East.A.f$, $\Domain{A.p^\omega}=\Domain{\Path_{k_1}.p^\omega}\subseteq \NonCausal{\Path_{k_1}}$ and therefore $\Path_{k_1}.p^{\omega}$ is a subgraph in \AlphaMax, as depicted in \RefFig{fig:infinitely-pumpable}, and there is an ultimately periodic assembly path in \AlphaMax.
\end{proof}


\section{Conclusion}
\label{sec:conclusion}

We showed that for every confluent (deterministic) temperature 1 system  \TAS, if the maximal assembly is infinite, then \AlphaMax contains an ultimately periodic assembly path.
In our proof we used notions of `left' and `right' regions of a bi-infinite path. 
Being in two dimensions, the Jordan curve theorem ensures that a bi-infinite path divides the plane in two regions, and hence $\IntegerSet^2$ is divided in two regions by a bi-infinite path.
In our case, the regions are connected subgraphs of $\IntegerSet^2$ and both regions contain the path itself.
The notions of left and right of a path are also used in \citep{meunier+woods17stoc,meunier+regnault+woods20stoc}, except in our case the paths are bi-infinite and are not necessarily part of \AlphaMax.
The other tool developed for the proof is the co-grow of two paths, which is a function that produces a free path that is a superposition of the two paths by taking the rightmost turns of the two paths (a similar notion of `right-priority' was used  in~\citep{meunier+woods17stoc,meunier+regnault+woods20stoc}). 
This tool provides a way to identify, or construct, an ultimately periodic path in \AlphaMax.
In order to co-grow two paths we relied on the confluence of the system.

The co-grow can be applied in any confluent two-dimensional system.
It may be of interest to extend this notion to systems that are not necessarily confluent, nor at temperature 1, or even in higher dimensions.
We believe that the presence of ultimately periodic assembly paths is not decidable in non-confluent systems, following the result of undecidability of growing infinite ribbons \citep{adleman+kari+kari+reishus+sosik09siam}.

Our result is a first step toward identifying the two basic structures that an infinite \AlphaMax is conjectured to have: a grid, or a finite union of combs~\citep{doty+patitz+summers11tcs}. These two structures were identified as main structures of \AlphaMax in confluent systems if the pumpability of paths is assumed~\citep{doty+patitz+summers11tcs}. We do not show the necessary pumping lemma, but we believe that the main result of~\RefTh{th:nice}, i.e., the merely existence of an ultimately periodic path, and the condition 2. of
~\RefLem{lem:nice} provides a tool for characterizing \AlphaMax through a design of a finite automaton-like structure (called `quipu')~\citep{durand-lose+hoogeboom+jonoska19arxiv}. Quipu is a specific automaton associated with a given \AlphaMax that contains a cycle for every ultimately periodic path.
The finiteness of the quipu and the fact that it is constructible would imply that a temperature 1 system \TAS cannot have universal computing power.

\paragraph{Acknowledgement.}
NJ  was (partially) supported by the grants NSF DMS-1800443/1764366 and the Southeast Center for Mathematics and Biology, an NSF-Simons Research Center for Mathematics of Complex Biological Systems, under National Science Foundation Grant No. DMS-1764406 and Simons Foundation Grant No. 594594.
We kindly acknowledge institutional support of the University of South Florida in 2017 to JDL and HJH whose visit initiated this work, and University of Orl\'eans visiting professorship grant to NJ in 2018.


\small

\bibliographystyle{plainnat}

\bibliography{bib_jdl-part1,bib_paper}

\clearpage
\appendix



\JDLvocabularyTableofsymbols


\end{document}
